\documentclass[12pt,a4paper]{article}

\usepackage{a4wide}
\usepackage{latexsym}
\usepackage{epsf}
\usepackage{amssymb}
\usepackage{graphicx}
\usepackage{cite}
\usepackage{bbm}
\usepackage{color}
\usepackage{amsmath,amssymb,amsthm}
\usepackage{pst-all}
\usepackage{psfrag}
\usepackage{verbatim}
\usepackage{hyperref}
\usepackage{subfigure}

\renewcommand{\d}{\textrm{d}}
\newcommand{\Real}{\textrm{I\!R}}

\newcommand{\e}{\textrm{e}}
\renewcommand{\d}{\textrm{d}}

\newcommand{\SU}{\mathop{\rm SU}}

\newcommand{\SO}{\mathop{\rm SO}}

\newcommand{\U}{\mathop{\rm {}U}}

\newcommand{\Sp}{\mathop{\rm {}Sp}}

\newcommand{\GL}{\mathop{\rm GL}}

\newcommand{\eq}[1]{\begin{equation}#1\end{equation}}
\newcommand{\spl}[1]{\begin{split}#1\end{split}}
\newcommand{\al}[1]{\begin{align}#1\end{align}}
\newcommand{\subeq}[1]{\begin{subequations}#1\end{subequations}}

\newcommand{\be}{\begin{equation}}
\newcommand{\ee}{\end{equation}}
\newcommand{\ba}{\begin{eqnarray}}
\newcommand{\ea}{\end{eqnarray}}
\newcommand{\nn}{\nonumber}
\newcommand{\lp}{\left(}
\newcommand{\rp}{\right)}
\renewcommand{\k}{\kappa}
\renewcommand{\a}{\alpha}
\renewcommand{\b}{\beta}

\newcommand{\s}{\sigma}
\newcommand{\w}{\wedge}

\newcommand{\Om}{\Omega}

\newcommand{\D}{\textrm{D}}

\newcommand{\F}{\mathcal{F}}

\begin{document}
\numberwithin{equation}{section}
\begin{flushright}
\small

UUITP-05/11\\
MAD-TH-10-04
\date \\
\normalsize
\end{flushright}

\begin{center}

\vspace{.4cm}

{\LARGE {\bf De Sitter hunting in a classical landscape}} \\

\vspace{1.1cm}

{\large Ulf H.\ Danielsson$^a$, Sheikh S.\ Haque$^b$, Paul Koerber$^c$,\\
\vspace{0.2cm} Gary Shiu$^{b,d}$, Thomas Van Riet$^a$ and Timm
Wrase$^{d,e}$} \footnote{ulf.danielsson \emph{at} physics.uu.se, haque
\emph{at} wisc.edu, koerber \emph{at} itf.fys.kuleuven.be, shiu
\emph{at} physics.wisc.edu, thomas.vanriet \emph{at} fysast.uu.se,
timm.wrase \emph{at} cornell.edu } \vspace{1cm}

$^a$ {\small\slshape Institutionen f\"{o}r Fysik och Astronomi,\\
Box 803, SE-751 08 Uppsala, Sweden}\\
\vspace{.3cm} $^b$ {\small\slshape Department of Physics,
University of Wisconsin,\\ Madison, WI 53706, USA}\\
\vspace{.3cm}
$^c$ {\small\slshape Instituut voor Theoretische Fysica, K.U.Leuven, \\Celestijnenlaan 200D, B-3001 Leuven, Belgium} \\
\vspace{.3cm}
$^d$ {\small\slshape Institute for Advanced Study, Hong Kong University of Science and Technology, \\
Hong Kong, People's Republic of China} \\
\vspace{.3cm}
$^e$ {\small\slshape Department of Physics, Cornell University,\\ Ithaca, NY 14853, USA} \\

\vspace{1cm}

{\bf Abstract} \end{center} {\small } We elaborate on the
construction of de Sitter solutions from IIA orientifolds of
$\SU(3)$-structure manifolds that solve the 10-dimensional equations
of motion at tree-level in the approximation of smeared sources.
First we classify geometries that are orbifolds of a group manifold
covering space which, upon the proper inclusion of O6 planes, can be
described within the framework of $\mathcal{N}=1$ supergravity in
4D. Then we scan systematically for de Sitter solutions, obtained as
critical points of an effective 4D potential. Apart from finding
many new solutions we emphasize the challenges in constructing
explicit classical de Sitter vacua, which have sofar not been met.
These challenges are interesting avenues for further research and
include finding solutions that are perturbatively stable, satisfy
charge and flux quantization, and have genuine localized (versus
smeared) orientifold sources. This paper intends to be
self-contained and pedagogical, and thus can serve as a guide to the
necessary technical tools required for this line of research. In an
appendix we explain how to study flux and charge quantization in the
presence of a non-trivial $H$-field using twisted homology.

\newpage

\pagestyle{plain}

\tableofcontents

\section{Introduction}

The increasing experimental support for an accelerating universe
also presents a great challenge for quantum gravity. The simplest
explanation for this cosmic puzzle is that we are living in a de
Sitter universe, i.e.\ a (metastable) vacuum  with a positive
cosmological constant. While the zero of the vacuum energy is
immaterial in the absence of gravity, and the cosmological constant
can be tuned at will classically, it is not so in the context of
quantum gravity. Thus, finding an explicit de Sitter vacuum which
matches with (and hopefully explains) observations is arguably a
holy grail for any candidate quantum theory of gravity, such as
string theory.

Like the search for the Holy Grail, the search for de Sitter vacua
in string theory has proven to be an elusive quest. Before taking
into account the full quantum corrections to the cosmological
constant, the construction of de Sitter vacua even at leading
order\footnote{One can consider de Sitter vacua that arise only
after non-perturbative effects are included, as is often assumed in
heterotic string theory or in the KKLT scenario\cite{Kachru:2003aw}.
However, the uplift to de Sitter space requires anti-branes
\cite{Kachru:2003aw} or stringy corrections
\cite{Balasubramanian:2004uy,Balasubramanian:2005zx} whose explicit
microscopic description may invoke similar issues encountered here.}
(i.e.\ within classical supergravity) is complicated by many additional issues,
such as moduli stabilization (and vacuum stability),
flux quantization, and a proper microscopic treatment of localized sources.
These issues, all  in one way or the other related to the fact that
string theory comes with extra dimensions of space,
may well turn out to give us
important leads for finding de Sitter vacua in a fundamental theory of gravity.

Particularly on the issue of moduli stabilization, the difficulty it
imposes on constructing de Sitter vacua is sharpened by several
no-go theorems. These theorems were presented in the language of 4D
effective field theories: by requiring the existence of a
positive-energy extremum of the potential in the dilaton and
breathing mode direction \cite{Hertzberg:2007wc, Silverstein:2007ac,
Caviezel:2008ik, Haque:2008jz, Caviezel:2008tf, Flauger:2008ad, Danielsson:2009ff,
Caviezel:2009tu, Wrase:2010ew}, one finds necessary conditions for
obtaining de Sitter vacua within classical supergravity. We refer to
de Sitter backgrounds obtained within supergravity, with the
possibility of localized sources and fluxes, but without introducing
stringy corrections or non-perturbative effects, as {\it classical
de Sitter solutions}. Attempts to construct actual classical de
Sitter solutions from 4D effective field theory were made
in\cite{Saltman:2004jh, Silverstein:2007ac, Haque:2008jz,
Flauger:2008ad, Caviezel:2008tf, deCarlos:2009fq, deCarlos:2009qm,
Caviezel:2009tu,  Dong:2010pm, Andriot:2010ju} (see also
\cite{Dibitetto:2010rg}).  Moreover, not many of such 4D solutions
were demonstrated to arise as full supergravity backgrounds,
especially those that break supersymmetry at the Kaluza-Klein scale
\cite{Saltman:2004jh, Silverstein:2007ac, Haque:2008jz,
Andriot:2010ju, Dong:2010pm}. This motivated one to construct
classical de Sitter solutions directly in 10 dimensions
\cite{Danielsson:2009ff, Danielsson:2010bc}.\footnote{Here, we
mention in passing that compactifications with non-geometric fluxes
seem to lead to stable de Sitter vacua
\cite{deCarlos:2009fq,deCarlos:2009qm} though it is not clear if a
supergravity approach is reliable for such constructions. Finally,
the aim of the present work is to search for de Sitter vacua from
{\it compactifications} of string theory. We refer the readers to
\cite{Gibbons:2001wy, Kallosh:2002gg, Kallosh:2001gr, Roest:2009tt,
Neupane:2010is} for some non-compact examples.} The work of
\cite{Danielsson:2010bc} explicitly mapped a solution obtained in
10D to a solution obtained in \cite{Caviezel:2008tf} using 4D
techniques. This solution has a simple and elegant form in the full
10D description and will be used in this paper to investigate charge
and flux quantization.

In this paper, we systematically explore the possibility of
constructing classical de Sitter vacua. Armed with the insights from
previous attempts, we consider Type IIA orientifolds of
$\SU(3)$-structure manifolds. In particular, we classify
$\SU(3)$-structure manifolds that are realized as orbifolds of a
group (or coset) manifold as the analysis is much simplified for
homogeneous spaces.\footnote{For this reason coset and group manifolds were already studied
early on, for instance in the context of the
Kaluza-Klein approach to unification. For reviews see
\cite{KKreview,zoup} and references therein. For other early work, e.g.\ in the context of heterotic string theory,
see \cite{lustcoset1,lustcoset2,hetcos3}.} We are particularly interested in
compactifications which do not break supersymmetry at the
Kaluza-Klein scale.\footnote{More precisely, this means we
consider compactifications that do not include explicit SUSY
breaking orientifold sources, so that SUSY is broken by the
(geometric) fluxes only.} The de Sitter solutions so obtained are
critical points of an effective 4D potential; supersymmetry is
broken spontaneously rather than explicitly and thus the de Sitter
backgrounds are under better analytic control. However, the on-shell
value for the volume modulus might be such that the KK scale is of
the same order as the flux-induced masses (which roughly give the
SUSY breaking scale). Then one has to worry about the consistency of
the effective potential. But since we restrict to left-invariant
modes on homogenous spaces, the dimensional reduction is
mathematically consistent and solutions of the four-dimensional
action lift to solutions of the ten-dimensional action
\cite{Cassani:2009ck}.

Moduli stabilization as well as flux and charge quantization are
subjects of concern in this paper, but we leave the challenge of finding
de Sitter backgrounds with account of the back-reaction of localized
sources for future work. As in the previous literature on the
subject, we will therefore look for supergravity solutions where the
O-planes are smeared over their transverse directions
\cite{Acharya:2006ne,Koerber:2007hd,Caviezel:2008ik,Blaback:2010sj,Blaback:2011}. One can think of the ``smeared"
approximation as solving the equations of motion in an average
sense, even though they are not satisfied pointwise in the internal
space \cite{Douglas:2010rt, Blaback:2010sj,Blaback:2011}.

An important obstacle for all the known dS solutions in these
compactifications is that \emph{none} sofar are perturbatively
stable. This is why one needs to increase the number of models, as
we do here, in the hope to find a perturbatively stable solution.
Unfortunately all new solutions we find in this paper have at least
one tachyonic mode in the left-invariant spectrum.

The paper is organized as follows. In section
\ref{4D10D}, we discuss the reduction of type IIA string theory on
$\SU(3)$ structure manifolds, providing both a 4D and a 10D
perspective. We emphasize the utility of the ``universal" ansatz in
the search for explicit solutions. In section \ref{Geometries}, we
classify homogenous geometries that are consistent with an
$\SU(3)$-structure. Our classification of group (and coset)
manifolds in this section subsumes all the previously found examples
in the literature and furthermore provides a unified, group
theoretical understanding of these previous works. Section \ref{sec:discrete}
discusses orbifold and orientifold projections and contains some explicit
examples of group spaces with $\SU(3)$-structure. We then apply the
results of the previous two sections to explore de Sitter solutions in
section \ref{deSitterSolutions}. In addition to solving the
equations of motion --- which are local conditions --- tadpole
cancelation and quantization of charges and fluxes impose additional,
global, constraints. We discuss these global constraints in section
\ref{quantization} using twisted homology \cite{twistedhom}. We explicitly quantize
the charge and fluxes of the solution on $\SU(2)\times \SU(2)$,
presented in \cite{Caviezel:2008tf, Danielsson:2010bc} and shall see
that this solution does not survive quantization. Therefore
quantization can serve as a strong criterium in de Sitter model
building. We end with some concluding remarks in section
\ref{Discussion}. Some details are relegated to the appendices.


\section{The 4D and 10D picture}\label{4D10D}

In this section we explain two alternative ways of searching for dS
vacua. First we discuss the reduction of the 10D type IIA theory on
an $\SU(3)$-structure manifold to four dimensions in subsection
\ref{4d}. This leads to a 4D effective theory with a scalar
potential $V$. One can then ask whether minima of $V$ exist for
which $V>0$. This approach is widely used in the literature but
certain questions can be easily addressed if one works directly with
the 10D equations of motion. We discuss this in subsection \ref{10d}
where we also list some advantages of this approach.

\subsection{Dimensional reduction and effective 4D theories}\label{4d}

Before one can reduce the action from ten to four dimensions one has
to decide which fields one keeps in the 4D theory. For
compactifications on an $\SU(3)$-holonomy manifold $M_6$ (i.e.\ when
$M_6$ is a $CY_3$ manifold) one normally expands the 10D fields in
the cohomology of $M_6$. Allowing the coefficients in this expansion
to depend on the four non-compact directions, they become 4D fields
in the effective 4D theory upon integrating the 10D action over
$M_6$. These fields are the lightest fields of an infinite tower of
Kaluza-Klein (KK) fields that result from the compactification.
Because these 4D scalar fields would be massless in the absence of
fluxes, we will refer to them as moduli (even when now they have
flux-induced masses). Restricting to these fields gives an effective
4D low-energy theory that is valid for energies below the KK-scale,
which is approximately given by the inverse length of the cycles in
$M_6$. Consistency requires that the flux-induced masses for the
scalar fields are smaller than the KK-scale which should generically
be true for sufficiently large $M_6$ \cite{DeWolfe:2005uu}.

When $M_6$ is a non-Ricci flat $\SU(3)$-structure manifold, it is
not clear anymore what the lightest fields are and how to obtain a
low-energy effective action
\cite{KashaniPoor:2006si,KashaniPoor:2007tr}. However, for the specific case of group
manifolds we can restrict to expansion forms that are
left-invariant under the group action. This leads to a 4D theory
that is a consistent truncation \cite{Cassani:2009ck}, i.e.\ a
solution to the 4D equations of motion will also be a solution to
the 10D equations of motion. This should not change when one
additionally does an orientifold projection. This was explicitly
confirmed in \cite{Danielsson:2010bc} where it was shown that the 4D dS solution of
\cite{Caviezel:2008tf} also solves the 10D equations of
motion. We will therefore restrict ourselves to group spaces and
expand the 10D fields and fluxes in left-invariant forms. Our
conventions for the forms are given in appendix
\ref{sec:Conventions}.

We now describe the 4D fields in the bosonic sector (the fermionic
fields are determined by $\mathcal{N}=1$ supersymmetry). On an
$\SU(3)$-structure manifold $M_6$ there exists a real two-form
$J=k^i Y^{(2-)}_i$ and a holomorphic three-form $\Omega$, which are
generalization of the K\"ahler form and holomorphic three-form that
exist on $CY_3$ manifolds. The volume of $M_6$ in string frame is
given by
\be
\text{vol}_6 = \frac16 \int J \w J \w J = \frac16 \k_{ijk} k^i k^j k^k,
\ee
and we use conventions in which $\Om \w \Om^* = \frac{4i}{3} J\w J\w J$.
In order to preserve only $\mathcal{N}=1$ supersymmetry in 4D we do an orientifold projection,
which contains an anti-holomorphic involution $\sigma$. This leads to O6-planes
that extend along the external four dimensions and wrap internal 3-cycles that
are the fixed loci of $\sigma$. The action of $\sigma$ on $J$ and $\Om$ is
\be
\s:\,\Om \rightarrow -\Om^*,\qquad \s: J \rightarrow -J,
\ee
so that we can expand $\Om = \Omega_R + i \Omega_I
=\mathcal{F}_K Y^{(3-)K} + i \mathcal{Z}^K Y^{(3+)}_K$, where
$\mathcal{F}_K$ and $\mathcal{Z}^K$ are real. Here the + and the - indicate
the parity under the orientifold involution. For more notation conventions about
our left-invariant forms see appendix \ref{sec:Conventions}.
Note that the
$\mathcal{F}_K$ are functions of the $\mathcal{Z}^K$ and therefore
not independent. The $\mathcal{Z}^K$ are projective coordinates so
that only $h^{2,1}$ of them are independent.

In addition to these fields we have in type IIA supergravity the
dilaton $\phi$, the Kalb-Ramond two-form $B$ and the RR three-form
$C_3$. Note that on a homogeneous manifold the RR form $C_1$ is not
compatible with a strict $\SU(3)$-structure, which requires the
absence of nowhere-vanishing one-forms. The 4D scalars arising from
the reduction of these fields can be grouped into two sets of
complex scalars \subeq{\al{
J_c &= J -i B= t^i Y^{(2-)}_i = (k^i -i b^i) Y^{(2-)}_i,\\
\Om_c &= \e^{-\phi} \text{Im}(\Om) + i C_3= N^K Y^{(3+)}_K = (u^K +
i c^K) Y^{(3+)}_K\, . }}
Note that $\e^{-\phi} \text{Im}(\Om)$ combines the dilaton with the
$\mathcal{Z}^K$ so that all the $h^{2,1}+1$ of the $u^K$ are independent
degrees of freedom.

The reduction of the $C_3$ field also leads to 4D
vector fields
\be
C_3 = A^\a Y^{(2+)}_\a,
\ee
so that we  have the 4D gauge group $U(1)^{h^{1,1}_+}$. These are the only
vector fields in the 4D theory since there are no
one-forms that could lead to vector fields coming from the metric and
$B$-field, and the $C_1$ field is odd under the orientifold
projection so that it is projected out. Besides these scalar and
vector fields and the 4D metric there are no further bosonic fields in the 4D
theory.

Similarly to the fields we can expand the fluxes  in terms of
left-invariant forms which have the correct parity under the
orientifold projection\footnote{The only external flux that preserves 4D Lorentz
symmetry is $F_4$-flux along the four space-time directions. We find it more convenient
to work with its dual $\hat{F}_6$.}
\be
H = h_K Y^{(3-)K}, \quad \hat{F}_0 = f_{(0)},\quad \hat{F}_2 =f_{(2)}^i Y^{(2-)}_i,
\quad \hat{F}_4 = f_{(4)i} Y^{(4+)i},\quad \hat{F}_6 = f_{(6)} Y^{(6-)}.
\ee
For a general $\SU(3)$-structure manifold the left-invariant forms are not
necessarily in cohomology and therefore they are not necessarily
closed under the exterior derivative $\d$. However, $\d$ maps
left-invariant forms into left-invariant forms so that we can write
\be
\d Y^{(2-)}_i = -r_{iK} Y^{(3-)K},\qquad \d Y^{(2+)}_\a = -\hat{r}_\a{}^K Y^{(3+)}_K.
\ee
Since $\int \d \lp Y^{(2\mp)} \w Y^{(3\pm)} \rp=0$, this implies
\be
\d Y^{(3+)}_K = -r_{iK} Y^{(4+)i}, \qquad \d Y^{(3-)K} = \hat{r}_\a{}^K Y^{(4-)\a}.
\ee
The matrices $r_{iK}$ and $\hat{r}_\a{}^K$ contain so called metric
fluxes that are T-dual to $H$-flux. On group manifolds with
$\SU(3)$-structure one always has six global left-invariant
one-forms $e^a, \, a=1,\ldots,6$ and one often defines the metric
fluxes $f^a{}_{bc}$ through $\d e^a=- \frac12 f^a{}_{bc} e^b \w
e^c$. The $r_{iK}$ and $\hat{r}_\a{}^K$ are then linear functions of
the $f^a{}_{bc}$.

The action of $\mathcal{N}=1$ supergravity in four dimensions\footnote{Except in section \ref{quantization} and appendix \ref{app:quant} we set $2 \pi \sqrt{\alpha'}=1$. The pre-factor $\tfrac{1}{2 \kappa_{10}^2}=\tfrac{1}{(2\pi)^7 (\alpha')^4}=2\pi$ in the usual 10D action can be removed by rescaling the action.} is
determined by the real K\"ahler potential $K$, the holomorphic
superpotential $W$ and the holomorphic gauge kinetic couplings
$f_{\a\b}$
\ba
S^{(4)} &=& - \int \left[ -\frac12 R \star_4 1 + K_{i\bar{\jmath}} \d t^i \w \star_4 \d \overline{t^j}+ K_{K\bar{L}} \d N^K \w \star_4 \d \overline{N^L} + V \star_4 1 \right.\\
&&\qquad \left. +\frac12 \text{Re}(f_{\a\b}) F^\a \w \star_4 F^\b +\frac12 \text{Im}(f_{\a\b}) F^\a \w F^\b \right],\nn
\ea
where $K_{i \bar{\jmath}} = \partial_{t^i} \overline{\partial_{t^j}} K$, $K^{i \bar{\jmath}}$
is its transposed inverse and similarly for the $N^K$. The
field strengths are $F^\a=\d A^\a$ and the 4D scalar
potential is
\be\label{eq:scalarpotential}
V = \e^K \lp K^{i\bar{\jmath}} \D_{t^i} W \overline{\D_{t^j} W} + K^{K\bar{L}} \D_{N^K} W \overline{\D_{N^L} W}- 3
|W|^2\rp +\frac12 {\lp \text{Re} f \rp^{-1}}^{\a\b} D_\a D_\b,
\ee
where the derivatives $\D_{t^i} W = \partial_{t^i} W + W \partial_{t^i}
K$ (and analogously for $\D_{N^K}$) should not be confused with the
D-terms $D_\a$ which are given by
\be\label{eq:Dterm}
D_\a = i (\delta_\a t^i \partial_{t^i} K +\delta_\a N^K \partial_{N^K} K) + i \frac{\delta_\a W}{W}.
\ee
Here $\lambda^\a \delta_\a t^i$ is the
variation of the field $t^i$ under the infinitesimal gauge
transformation $A^\a \rightarrow A^\a + \d \lambda^\a$ and
similarly for $N^K$ and $W$. Equation \eqref{eq:Dterm} is only valid
for $W \neq 0$ but for the reductions we consider here (and
generalizations thereof) one can show that $\delta_\a W$ always
vanishes \cite{Ihl:2007ah,Robbins:2007yv}. This means that the last term in
\eqref{eq:Dterm} vanishes and equation \eqref{eq:Dterm} without the last term is
true even for $W=0$.

The explicit forms of $K$, $W$, $f_{\a\b}$ and $D_\a$ for  the
reduction of type IIA supergravity on $\SU(3)$-structure spaces have
been derived in many papers \cite{Derendinger:2004jn, Grimm:2004ua,
Villadoro:2005cu, DeWolfe:2005uu, House:2005yc, Camara:2005dc, Benmachiche:2006df, Ihl:2006pp,
Ihl:2007ah, Robbins:2007yv, Villadoro:2007tb}
and the results are \subeq{\label{eq:KWfD}\al{
K &= -2 \, \text{ln} \lp -i \int \e^{-2\phi} \Om \w \Om^* \rp -\text{ln} \lp \frac43\int J \w J \w J \rp = 4 \phi_4 - \text{ln} \lp 8 \text{vol}_6\rp,\;\;\\
\sqrt{2} W &= \int \lp \Om_c \w (-i H + \d J_c) + \e^{i J_c} \w \hat{F} \rp \nn\\
&= -N^K (i h_K + r_{iK} t^i) + f_{(6)} +i f_{(4)i} t^i - \frac12 \k_{ijk} f_{(2)}^i t^j t^k - \frac{i}{6} f_{(0)} \kappa_{ijk} t^i t^j t^k,\\
f_{\a\b} &= - \hat{\k}_{i \a\b} t^i,\\
D_\a &= \frac{\e^{\phi_4}}{\sqrt{2 \text{vol}_6}} \hat{r}_\a^K \F_K,
}} where $\phi_4$ is the 4D dilaton defined by $\e^{-\phi_4}=
\e^{-\phi} \sqrt{\text{vol}_6}$ and $\hat{F}=\hat{F}_0 + \hat{F}_2 +
\hat{F}_4 + \hat{F}_6$ is the sum of the RR fluxes.

There is one 10D equation of motion, the $\hat{F}_2$ Bianchi
identity, that is not taken into account in this 4D analysis. It
reads $\d \hat{F}_2 + \hat{F}_0 H = -j^{(\delta)}$, where
$j^{(\delta)}$ is related to the Poincar\'e dual of the submanifold wrapped
by the O6-planes (cf.\ section \ref{sec:discrete} and appendix \ref{app:quant}). In this paper we will
follow the literature and only solve the integrated equation $-r_{iK}
f_{(2)}^i Y^{(3-)K} + f_{(0)} h_K Y^{(3-)K} + j = 0$, where the delta-functions
in $j^{(\delta)}$ are replaced by one in $j$.\footnote{Since the integration essentially replaces
delta-functions with 1, solving integrated equations of motion is
often called 'smearing the sources' in the literature.} The reason
for this approximation is that, to our knowledge, there are no known
solutions for (fully back-reacted) intersecting O-planes since the
equations of motion with the delta-functions are very hard to solve.

For any given compactification space it is straightforward to solve
the F- and D-term equations \be \D_{t^i} W = \D_{N^K} W =0,\qquad
D_\a =0, \ee and to find supersymmetric AdS or Minkowski vacua. For
AdS vacua it is generically possible to stabilize all moduli except
when $\text{rank}(r_{iK}) < h^{2,1}$ in which case
$h^{2,1}-\text{rank}(r_{iK})$ $C_3$ axions remain flat directions
\cite{Ihl:2007ah}. Note that this is not really a problem since the
moduli space of the axions is compact. In supersymmetric Minkowski
vacua it is not possible to stabilize all geometric moduli
\cite{Micu:2007rd, Ihl:2007ah}.

Since all geometric moduli are stabilized in AdS vacua we know that
the scalar potential depends on all geometric moduli and it is
therefore sensible to ask whether the scalar potential allows for dS
vacua as well. It is straightforward but tedious to calculate the
explicit scalar potential \eqref{eq:scalarpotential} using
\eqref{eq:KWfD} (see equation (2.34) in \cite{Flauger:2008ad}). In
order to find dS vacua one has to minimize this scalar potential
with respect to all the moduli which is technically hard and can
normally only be done numerically and only for $\SU(3)$-structure
manifolds with very few scalar fields. The authors of
\cite{Hertzberg:2007ke} scanned numerically the scalar potentials of
the three models \cite{Villadoro:2005cu, DeWolfe:2005uu,Ihl:2006pp}
for regions that allow for slow-roll inflation or dS vacua, but did
not find any. This called for a different approach. In
\cite{Hertzberg:2007wc} the authors derived a no-go theorem that
forbids slow-roll inflation and dS vacua in type IIA flux
compactifications on $CY_3$ manifolds with $O6$-planes. However,
this no-go theorem can be circumvented in compactifications on
$\SU(3)$-structure manifolds that are not Ricci-flat, but instead
negatively curved. This is the framework we will explore.

\subsection{The ``universal'' ansatz in 10D}\label{10d}
4D effective field theories correctly describe fluctuations
around a string (or supergravity) solution, whenever the 4D solution
can be lifted to a full solution of the 10D equations of motion, or
at least to {\it approximate} solutions (in the sense described below).
If every solution can be lifted to a full 10D solution, the
reduction is said to be \emph{mathematically} consistent. Even when
this requirement is not satisfied, the reduction can still be
\emph{physically} consistent in the sense that while it does not
fully describe the 10D physics, the effects of the heavy fields, the ones
we ignore, are not relevant at low energies. This is what we meant by
approximate 10D solutions. As an example, reductions on
generic Calabi--Yau spaces have not proven to be mathematically consistent
but are expected to be physically consistent when the internal volume is
large enough such that we can ignore the KK modes. As
explained above, for internal spaces with enough symmetries we
expect dimensional reduction can be fully mathematically consistent
if we restrict to the left-invariant modes and work in the limit
that the sources are smeared.  In that sense we should not be
bothered with the question whether we are solving the 10D equations of
motion, although we do not always have a separation of scales in our
models. However as explained in \cite{Danielsson:2009ff,
Danielsson:2010bc} it is interesting to work in the 10D picture
because of several reasons: 1) Flux and charge quantization require
us to know the 10D solution; 2) Computing the back-reaction of the sources,
i.e.\ going beyond the smeared limit, is
something that requires the 10D picture as well; 3) In ten
dimensions one can establish solutions to the equations of motion
without fully specifying the internal manifold.

Let us elaborate on point 3. Points 1 and 2 are discussed
later in this paper. In order to find a de Sitter minimum in the
4D picture one has to minimize the scalar potential with
respect to the moduli. This requires an explicit choice of manifold
and that makes the procedure model dependent. On the other hand,
when one solves 10D equations of motion it becomes clear
that one can find solutions by just specifying properties of the
internal manifold, but without fixing it entirely. This is for
instance the case with SUSY AdS vacua in IIA from $\SU(3)$-structure
manifolds \cite{Lust:2004ig} (and also non-SUSY AdS vacua, see
\cite{Koerber:2010rn}). Therefore one constructs solutions for
\emph{classes} of manifolds that just obey the given properties. A
downside of this approach is that it is not possible to compute the
mass spectrum since that requires the explicit choice of the
manifold. We will refer to such solutions as \emph{universal}
solutions. The idea of universal solutions was applied to the case
of dS vacua in \cite{Danielsson:2009ff} and
\cite{Danielsson:2010bc}. We will now review this approach and
generalize it a bit.

So we start with an orientifold of a general $\SU(3)$-structure
manifold. As in the previous papers on the topic we make the
simplification that the $\SU(3)$-structure is half-flat (see
appendix \ref{halfflat}). In the 4D picture this restriction
corresponds to the assumption that there are no non-zero D-terms
(cf.\ appendix A of \cite{Ihl:2007ah}). It would be interesting to
extend this 10D approach to manifolds that are not half-flat.

A half-flat manifold possesses a set of canonical forms, which we
call universal forms and they are given by the (would-be) real
K\"ahler form $J$ and the (would be) holomorphic complex three-form
$\Omega$, and the torsion classes $W_1, W_2, W_3$
\begin{equation}
\text{universal forms:}\qquad \Bigl\{ J, \Omega, W_1, W_2,
W_3\Bigr\}\,.
\end{equation}
These then serve as natural expansion forms for the fluxes. Hence a
general ansatz for a solution could be given by
\subeq{\label{fluxansatz}\al{
e^{\Phi} \hat{F}_0 & = f_1 \, , \\
e^{\Phi} \hat{F}_2 & = f_2 J + f_3 \hat{W}_2 \, , \\
e^{\Phi} \hat{F}_4 & = f_4 J \wedge J + f_5 \hat{W}_2 \wedge J \, , \\
e^{\Phi} \hat{F}_6 & = f_6 \text{vol}_6 \, , \\
H & = f_7 \Omega_R + f_8 \hat{W}_3 \, , \\
j & = j_1 \Omega_R + j_2 \hat{W}_3 \, . }} where the fluxes are
decomposed as follows: \eq{ F = \hat{F} + \text{vol}_4 \wedge
\tilde{F} \, . } The fluxes $\hat{F}$ and $\tilde{F}$ have only
components in the internal dimensions.  We furthermore used the
notation of \cite{Danielsson:2010bc} where
$\hat{W}^i=(\sqrt{|W^i|^2})^{-1}\,W^i$ \footnote{We define the
square of a $p$-form as $A_p^2=\frac{1}{p!}A_{i_1\ldots
i_p}A^{i_1\ldots i_p}$.}. This ansatz is consistent with the
orientifold involutions for supersymmetrically embedded orientifold
planes \cite{Koerber:2007hd}. In order to check for which
coefficients $f_1,\ldots, f_8$ and $j_1, j_2$ we have a solution we
need the expression for the Ricci tensor as demanded by the Einstein
equations. The Ricci tensor for a general $\SU(3)$-structure
manifold has been established in \cite{bedulli-2007-4} and is
presented in appendix \ref{halfflat}. The relevant property is that
it is given in terms of the universal forms. It is for this reason
that a universal ansatz (where the fluxes and sources are given by
universal forms) is sensible, since the Einstein equation forces the
energy-momentum tensor to be made from universal forms. However, to
our surprise, most de Sitter solutions in the models we consider
below are not universal. This implies that there must be non-trivial
cancelations of the non-universal flux pieces in the energy-momentum
tensor.

It turns out that in order to find solutions different from the SUSY
AdS solutions, one needs to impose constraints on the universal
forms. These constraints are such that the equations of motion
imply fewer constraints and therefore make possible the existence of
new solutions. These constraints are
\subeq{\label{tclprop}\al{
\d \hat{W}_2 & = c_1 \Omega_R + d_1 \hat{W}_3 \label{first}\, , \\
\hat{W}_2 \wedge \hat{W}_2 & = c_2 J \wedge J + d_2 \hat{W}_2 \wedge
J \, , \label{W2W2decomp}\\
\d \star_6 \hat{W}_3 & = c_5 J \wedge J + c_3 \hat{W}_2 \wedge J \, , \\
\frac{1}{2} (\hat{W}_{3\,ikl} \hat{W}_{3\, j}{}^{kl})^+ &=  d_4
J_{ik} \hat{W}_{2}{}^k{}_j \, . }}
where
\subeq{\al{ & c_1 = -
\frac{w_2}{4}, \quad c_2 = - \frac{1}{3!}, \quad c_3  = - d_1, \quad
c_4 = \frac{1}{2}, \quad c_5 = \frac{w_3}{3!} \\ & d_2 = - \star_6
(\hat{W}_2 \wedge \hat{W}_2 \wedge \hat{W}_2) \,, }}
and
\be
w_2 = \sqrt{W_2^2}, \quad w_3 = \sqrt{W_3^2}.
\ee

It is then straightforward to put the ansatz into the IIA equations
of motion (see \cite{Danielsson:2010bc}) to obtain the algebraic
equations for the flux parameters. These are very lengthy
expressions and we therefore present them in appendix \ref{ap:universaleoms}. It is very
non-trivial to find the general solution to these algebraic
equations but many solutions have nonetheless been found.

Let us review these solutions
\begin{itemize}
\item The SUSY AdS solutions necessarily have $W_3=0$ and they require us to enforce the
first constraint  in equations (\ref{tclprop})
\cite{Lust:2004ig, Koerber:2008rx}.
\item Non-SUSY AdS solutions can be found when $W_3=0$ when
we also enforce the second constraint in equations (\ref{tclprop})
\cite{Danielsson:2009ff, Koerber:2010rn}.
\item De Sitter solutions can be found under the same
circumstances as the above non-SUSY AdS vacua
\cite{Danielsson:2009ff}, however no explicit geometry has been
found that satisfies the parameter windows that gives these dS
solutions, as opposed to the AdS solutions. \item Universal
solutions with non-zero $W_3$ have been investigated in
\cite{Danielsson:2010bc} but with the simplification that $W_2=0$.
In that case AdS, Minkowski and dS solutions are possible when we
enforce the third and fourth constraint in equations (\ref{tclprop})
with the choice $d_4=0$. In fact one extra constraint was necessary,
namely $Q_1(\hat{W}_3, \hat{W}_3)\propto Q_2(\hat{W}_3,
\hat{W}_3)\propto (\hat{W}_3)_{2,1}$, where we refer to appendix
\ref{halfflat} for the definitions of $Q_1$ and $Q_2$.
Interestingly, there exists at least one explicit geometry that satisfies the
conditions for these universal dS solutions, namely $\SU(2)\times \SU(2)$ as was shown in
\cite{Danielsson:2010bc}.
\end{itemize}

\section{Classification of geometries}\label{Geometries}
\subsection{Homogenous $\SU(3)$-structures}
We want to classify homogeneous geometries that are consistent with
an $\SU(3)$-structure that is invariant under the left acting
isometries. The covering space of a homogenous manifold is
necessarily a coset $G/H$ (or a group when $H=1$). Furthermore one
can show that $H \subseteq \SU(3)$ if we want the coset to allow an
invariant $\SU(3)$-structure \cite{Koerber:2008rx}\footnote{If $H$
is not within $\SU(3)$ there could still be other non-G-invariant
$\SU(3)$-structures, but they are outside this particular coset
description. It could be (but by no means necessarily so) that they
are described by another coset description, because sometimes the
same manifold can have different coset descriptions,
e.g.~$\mathbb{CP}^3=\Sp(2)/S(\U(2)\times\U(1))=
\SU(4)/S(\U(3)\times\U(1))$. In this case there is a
$\Sp(2)$-invariant homogeneous $\SU(3)$-structure, but no
$\SU(4)$-invariant one.}.

The requirement of homogeneity and an invariant $\SU(3)$-structure
are strong constraints and the problem of classifying the spaces
turns into an algebraic problem that involves two steps. Firstly,
one has to use group theory to classify the possible covering spaces
$G/H$. Secondly, one has to classify the possible discrete subgroups
$L \subset G$ that one can use to create another space $G/H/L$. For
instance, when the covering space is non-compact we need to find a
discrete subgroup that can render the manifold compact after
division. Furthermore we need discrete subgroups that contain
involutions for the orientifold action.

\begin{figure}[here]
\begin{center}
\includegraphics[height=3.5in,width=5in]{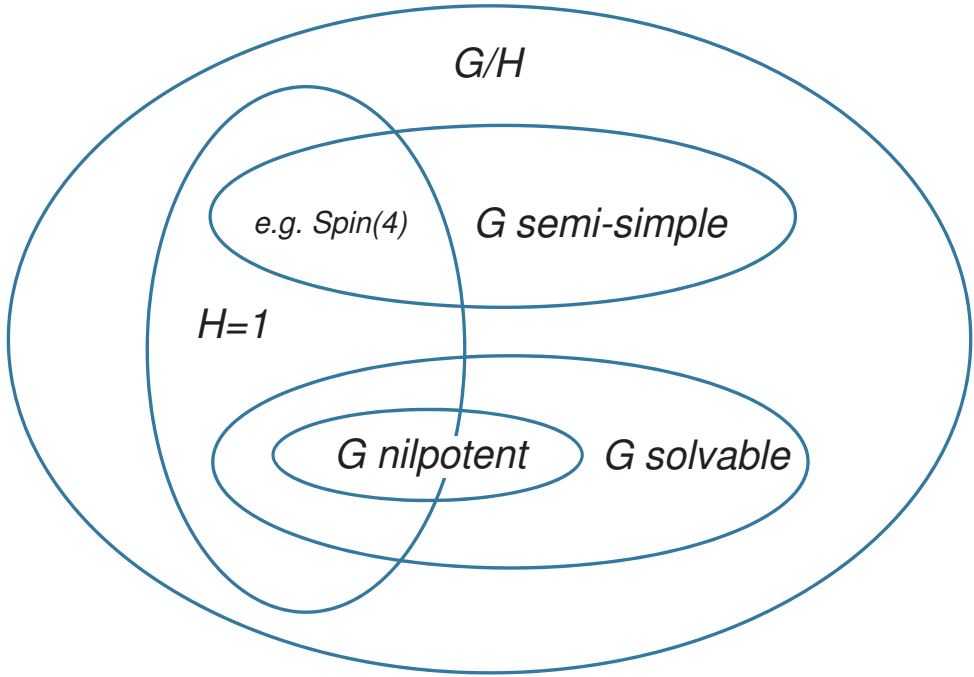}
\end{center}
\caption{The homogeneous $\SU(3)$-structure spaces $G/H$. The
isotropy group is necessarily contained in $\SU(3)$ for a
$G$-invariant $\SU(3)$-structure.} \label{fig:cosets}
\end{figure}

In figure \ref{fig:cosets} we sketch the classification of (the
covering spaces of) homogeneous $\SU(3)$-structure manifolds.

The picture shows the separate subset of cosets that are group
spaces ($H=\{1\}$). \footnote{Groups and cosets are a trivial kind
of $\SU(3)$-structure since the structure group of the frame bundle
is trivial. In this paper we consider orbifolds therefor such that
the frame bundle is not trivial anymore and we have a genuine
$\SU(3)$-structure, after blowing up the orbifold singularities.}
Lie groups can be classified using the properties of the associated
Lie algebra that can range from being semi-simple to the opposite,
being nilpotent. The solvable cases are somewhere in between.
Especially the nilpotent examples have received much attention since
they are the covering space of compact spaces obtained from
T-duality of a torus with $H$-flux. They are called twisted tori.
But the name twisted tori is sometimes also given to the solvable
cases and even all group spaces.

Let us discuss those cases that have already made an appearance in the
literature. Reference \cite{Koerber:2008rx} classified those cosets
that have an isometry group $G$ which is a compact semi-simple group
or the product thereof with U(1)-factors. This also includes the
space Spin(4) $= \SU(2) \times \SU(2)$, which we have depicted in
the intersection of groups and cosets with semi-simple isometry
groups.

The class of cosets in \cite{Koerber:2008rx} is not the most general
class of cosets with non-trivial isotropy because one can consider
spaces with non-compact $G$ or with $G$ generically non-semi-simple.
Examples of this sort have not appeared in the literature to our
knowledge. We will not deal with those examples here, but they might
offer an interesting class of coset geometries to investigate de
Sitter solutions. Reference \cite{Flauger:2008ad} considered all
possible ``metric fluxes'' consistent with the symmetries of all the
abelian orbifold groups of the six-torus. However they did not
provide any further description of these spaces like for example the
Lie algebra of symmetries. In contrast reference
\cite{Grana:2006kf}\footnote{Later this has been significantly been
improved in \cite{Andriot:2010ju}.} made a less complete list
(solvmanifolds and nilmanifolds), but has a partial description of
the geometries. Below we present a full (and simple) classification
of covering spaces $G$ using group theory, which allows us to have
an algebraic understanding of the various metric fluxes. We also
find more possibilities for orientifolds than the ones given in
\cite{Grana:2006kf}.

\subsection{Group manifold geometry and geometric moduli}
We recall the basic concepts of the geometry of a group manifold.
Although this has been done in many places before, we repeat this
here such that the paper is self-contained.

On a group manifold, $G$, one can define a co-frame of
left-invariant forms, called the Maurer--Cartan forms, as follows,
\begin{equation}
g^{-1}\d g =  e^a T_a\,,
\end{equation}
where the $T^a$ are the generators of the Lie algebra associated
to $G$ and denoted $\frak{g}$. The Maurer--Cartan forms obey the
relations
\begin{equation}
\d e^a =-\tfrac{1}{2}f^a{}_{bc}e^b\wedge e^c\,,
\end{equation}
where the $f$'s are the structure constants of $\frak{g}$. The
conditions for a set of $f^a{}_{bc}$ to describe a Lie algebra are
\begin{align}
&f^{a}{}_{bc}=-f^a{}_{cb}\,,\qquad f^{a}{}_{b[c} f^b{}_{de]}=0\,.
\end{align}

It can be shown that only unipotent groups, i.e.\ groups with traceless
structure constants,
\begin{equation}
f^a{}_{a b}=0 \,,\quad\text{for all}\,\, b\,\,\,\text{(sum over $a$
implied)}\,,
\end{equation}
can be made compact. To be more precise, only unipotent groups $G$
can have a discrete subgroup, $L$, acting without fixed points, such
that $G/L$ is a compact space. If the group is already compact $L$
can of course taken to be trivial. Unipotence, however, is not a
sufficient condition to establish the existence of such a discrete
subgroup, see e.g.~\cite{Andriot:2010ju, Grana:2006kf} and
references therein.

In terms of the Maurer--Cartan one-forms, $e^a=e^a_{\mu}\d x^{\mu}$, we
can introduce a metric on the group manifold,
\begin{equation}
\d s^2 = \mathcal{M}_{ab}e^a\otimes e^b\,,
\end{equation}
where $\mathcal{M}$ is constant, symmetric and positive definite.
The Ricci tensor is then given by (see e.g.~\cite{cosetreview1})
\begin{equation}
R_{ab}=\tfrac{1}{4}f_{acd}f_{b}{}^{cd}-\tfrac{1}{2}f^{c}{}_{da}f^d{}_{cb}-\tfrac{1}{2}f_{cda}f^{cd}{}_{b}\, ,
\end{equation}
where we lower and raise indices using the metric $\mathcal{M}$ and its inverse.
Furthermore we restricted to algebras for which $f^a{}_{ab}=0$, (sum over $a$
implied), since we are not interested in non-compact models. The Ricci scalar then reads
\begin{equation}\label{Ricciscalar}
R=-\tfrac{1}{4}f_{abc}f^{abc}-\tfrac{1}{2}f_{cab}f^{acb}\,.
\end{equation}

The matrix $\mathcal{M}$  parameterizes that part of the group
manifold moduli space that is concerned with metric deviations along
left-invariant directions. This moduli space is
$\GL(n,\Real)/\SO(n)$, as can be seen from the fact that the matrix
$\cal{M}$ transforms under $GL(n,\Real)$-matrices, but is invariant
under an $\SO(n)$-group. When we consider orbifold and orientifold
symmetries we put further restrictions on $\mathcal{M}$ such that
the geometric moduli space gets truncated.

\subsection{All unipotent real six-dimensional Lie algebras}

Here, we would like to classify all six-dimensional real Lie algebras. The
classification is already done in the literature \cite{Turkowski},
but in our case the classification is simplified because we
restrict to unipotent algebras.

Lie algebras are classified depending on whether or not they have ideals.
We remind ourselves that an ideal $\mathfrak{i}$ of an algebra
$\mathfrak{g}$ is a subalgebra with the following property:
$[\mathfrak{g}, \mathfrak{i}] \subseteq \mathfrak{i}$. On the one
end of the collection of Lie algebras one has \emph{simple Lie}
algebras. They are defined as the Lie algebras that have no proper
ideal. Next to those come the \emph{semi-simple} Lie algebras which
have no proper abelian ideals. One can show that they are direct
sums of simple Lie algebras. The definition of a semi-simple Lie
algebra turns out to be equivalent to having a non-degenerate Killing form
$C$, where $C_{ab}=f^c{}_{da} f^d{}_{cb}$. On the other end of the spectrum
of Lie algebras are those that have big ideals. One way to built
ideals is by taking commutators. Consider the following sets of
subalgebras of $\mathfrak{g}$
\begin{equation}
\mathfrak{g}^n = [\mathfrak{g}^{n-1}, \mathfrak{g}^{n-1}]\,,\qquad
\mathfrak{g}_n = [\mathfrak{g}, \mathfrak{g}_{n-1}]\,,
\end{equation}
where $\mathfrak{g}^0=\mathfrak{g}_0=\mathfrak{g}$. Both the
upper-derived series, $\mathfrak{g}^n$, and the lower-derived
series, $\mathfrak{g}_n$, are obviously ideals. A \emph{solvable}
Lie algebra is a Lie algebra for which the upper-derived series
vanishes at some point and for a \emph{nilpotent} Lie algebra the
lower-derived series vanishes at some point. It is easy to show that
$\mathfrak{g}^n \subseteq \mathfrak{g}_n$, from which follows that
a nilpotent algebra is also solvable. One can prove that the nilpotent Lie
algebras form the building blocks for the solvable algebras by
taking semi-direct products, whereas the solvable and simple  Lie
algebras are the building blocks for all remaining Lie algebras,
again by taking semi-direct products. The latter is the content of
\emph{Levi's theorem}, which states that any Lie algebra
$\mathfrak{g}$ is the semi-direct product of a semi-simple algebra
$\mathfrak{s}$ with the largest solvable ideal $\frak{r}$
\begin{equation}
\mathfrak{g}=\mathfrak{s}\ltimes \frak{r}\,.
\end{equation}
The largest solvable ideal, $\frak{r}$, can be shown to be
unique and is called the radical. Using this theorem the problem
boils down to classifying semi-simple and solvable algebras of
dimension $d\leq 6$ and the possible semi-direct products between
the $m$-dimensional semi-simple algebra $\frak{s}$ and the
$n$-dimensional radical $\frak{r}$, where $m+n=6$.

Semi-simple and nilpotent algebras are always unipotent, but
unipotence has to be verified for solvable groups. Note that for a
general Lie algebra to be unipotent, the radical itself has to be
unipotent. Indeed, using indices $a',b'$ to run over $\frak{s}$
and $a'',b''$ to run over $\frak{r}$ we find
\begin{equation}
 0 = f^a{}_{ab''} =
f^{a'}{}_{a'b''} + f^{a''}{}_{a''b''} = f^{a''}{}_{a''b''} \,,
\end{equation}
since all $f^{a'}{}_{a'b''}=0$, as $\frak{r}$ is an ideal.

The classification of the semi-simple algebras of dimension $d\leq
6$ is straightforward and one can find six examples
\begin{itemize}
\item $\frak{so}(p,q)$ with $p+q=4$ \,. \item
$\frak{so}(3)\times\frak{so}(2,1)$\,. \item $\frak{so}(p,q)$ with
$p+q=3$\,.
\end{itemize}
Note that $\frak{so}(4)=\frak{so}(3)^2$ and
$\frak{so}(2,2)=\frak{so}(2,1)^2$. So, there are four
six-dimensional semi-simple algebras and only the compact case
$\frak{so}(4)$ has so far been used in the flux literature. For the
other three cases, it is not clear whether there are points in
the moduli space of these groups that allow a discrete subgroup
$L$ that makes $G/L$ a smooth compact space.

To find the other six-dimensional group spaces using Levi's theorem
we deduce that we need to know all six- and three-dimensional
solvable algebras. The six-dimensional unipotent solvable algebras
are the ones that describe the solvmanifolds and their
classification appears in e.g.~\cite{Grana:2006kf}\footnote{The
classification of solvable algebras in \cite{Grana:2006kf} is not
complete. There are some non-algebraic examples that are not given.
For a complete list we refer to \cite{Turkowskisolv} or
\cite{Andriot:2010ju}.}. We will not repeat those algebras here but
refer to tables 4 and 5 in \cite{Grana:2006kf}). The nilpotent ones
have the nice feature that the associated group space $G$ always
allows a discrete subgroup $L$ such that $G/L$ is a smooth compact
space, called nilmanifolds.

Let us now focus on the remaining algebras that have a
three-dimensional radical. This implies we have to find all
solvable unipotent three-dimensional Lie algebras first. This is
straightforward since they are contained in the list of
six-dimensional unipotent solvable algebras as those which have a
$\frak{u}(1)^3$-factor. Alternatively one can simply use the known
classification of three-dimensional Lie algebras (a.k.a. the
Bianchi classification, see e.g.~\cite{Roest:2004pk}) and check
which are unipotent and solvable. There are only four of these and
given explicitly by
\begin{itemize}
\item $\frak{u}(1)^3$\,, \item Heisenberg: \qquad $[r_1,
r_2]=r_3$\,.\item $\frak{iso}(2)$: $\qquad [r_1,
r_2]=r_3\,,\qquad [r_3, r_1]=r_2$\,,\item
$\frak{iso}(1,1)$: $\qquad [r_1,
r_2]=r_3\,,\qquad [r_3, r_1]=-r_2$\,,
\end{itemize}
where $r_1, r_2, r_3$ represent the generators and we only
presented the non-zero commutators.

The next step is to consider the possible semi-direct products of
these four solvable algebras with the two three-dimensional simple
algebras $\frak{so}(3)$ and $\frak{so}(2,1)$. Let us recall the
definition of a semi-direct product: a semi-direct product
$\mathfrak{s}\ltimes \frak{r}$ is the direct product as
vector spaces equipped with a Lie bracket
\begin{equation}
[(s_1, r_1), (s_2, r_2)]_{\mathfrak{s}\ltimes \frak{r}}=
([s_1, s_2]_{\mathfrak{s}},[r_1, r_2]_{\frak{r}} +
\rho(s_1)r_2-\rho(s_2)r_1) \,,
\end{equation}
for all $s_1, s_2 \in \mathfrak{s}$ and $r_1, r_2 \in
\frak{r}$. Here $\rho$ is a Lie algebra homomorphism of
$\mathfrak{s}$ into the derivative operators on $\frak{r}$.
To be explicit, this means
\subeq{\begin{align}
\rho([s_1,s_2])& =\rho(s_1)\rho(s_2)-\rho(s_2)\rho(s_1) \qquad \text{(Lie-algebra homomorphism)} \, , \\
\rho(s)([r_1,r_2])&=[\rho(s)(r_1),r_2]+[r_1,\rho(s)(r_2)] \qquad
\text{(derivative operator)}\label{condder} \,.
\end{align}
} Both properties are needed to assure the Jacobi identity of the
semi-direct product. Since $\rho$ is a representation of $\frak{s}$
in $\frak{r}$ it should be possible to classify the different
semi-direct products $\rho$ by the representation theory of
$\frak{s}$.

The possible representations of $\frak{so}(3)$ and
$\frak{so}(2,1)$ must be three-dimensional. So we have
\begin{equation}
\textbf{1} \oplus \textbf{1} \oplus \textbf{1}\,,\qquad \textbf{1}
\oplus \textbf{2}\,,\qquad  \textbf{3}\,.
\end{equation}
The representation $\textbf{3}$ for $\frak{so}(3)$ is
\begin{equation}
M_1=\begin{pmatrix} 0 & 1 & 0\\
-1& 0 & 0 \\
0 & 0 & 0\end{pmatrix}\,,\qquad M_2=\begin{pmatrix} 0 & 0 & 1\\
0 & 0 & 0 \\
-1& 0&0\end{pmatrix}\,,\qquad M_3=\begin{pmatrix} 0 & 0 & 0\\
0 &0 &1 \\
0&-1&0\end{pmatrix}\,.
\end{equation}
The representation $\textbf{3}$ for $\frak{so}(2,1)$ is
\begin{equation}
M_1=\begin{pmatrix} 0 & 1 & 0\\
1 &0 &0 \\
0&0&0\end{pmatrix}\,,\qquad M_2=\begin{pmatrix} 0 & 0 & 1\\
0 &0 &0 \\
1&0&0\end{pmatrix}\,,\qquad M_3=\begin{pmatrix} 0 & 0 & 0\\
0 &0 &1 \\
0&-1&0\end{pmatrix}\,.
\end{equation}
The representation $\textbf{2}$ for $\frak{so}(3)$ is given by the
Pauli matrices
\begin{equation}
M_1=\begin{pmatrix} 0 & 1 \\
1 &0
\end{pmatrix}\,,\qquad M_2=\begin{pmatrix} 0 & i \\
-i &0
\end{pmatrix}\,,\qquad M_3=\begin{pmatrix} 1 & 0 \\
0 &-1
\end{pmatrix}\,.
\end{equation}
Since the Pauli matrices contain at least one complex generator
(in any basis) we cannot build a real six-dimensional
Lie-algebra from this representation of the semi-direct product.
So the $\textbf{1} \oplus \textbf{2}$ is discarded for
$\frak{s}=\frak{so}(3)$. Through the Weyl unitary trick we find
the two-dimensional representation of $\frak{so}(2,1)$ to be real:
\begin{equation}
M_1=\begin{pmatrix} 0 & 1 \\
1 &0
\end{pmatrix}\,,\qquad M_2=\begin{pmatrix} 0 & 1 \\
-1 &0
\end{pmatrix}\,,\qquad M_3=\begin{pmatrix} 1 & 0 \\
0 &-1
\end{pmatrix}\,.
\end{equation}
Hence, the $\textbf{1} \oplus \textbf{2}$ is still possible for
$\frak{s}=\frak{so}(2,1)$.

Having recalled the various representations we need to go from the
representation $\rho$ to the construction of the six-dimensional
Lie algebra, that is, the construction of the structure constants.
For that, consider the bracket
\begin{equation}
[s_a, r_i]=f^j{}_{ai} r_j\equiv [M_a]^j_i r_j\,,
\end{equation}
The three matrices $M_a$ make
up the representation of the simple part $\frak{s}=(\frak{so}(3)$ or
$\frak{s}=\frak{so}(2,1)$).

For the $\textbf{1}$ representation we have $M_a=0$, hence the
six-dimensional algebra is the direct product of the solvable and
the simple part. Further constraints come from the demand that
$\rho$ is a derivation when $\rho$ is non-trivial (i.e.\ $\rho \neq
\textbf{1} \oplus \textbf{1} \oplus \textbf{1}$). We analyze this
below, where we denote by $\rho$ the representation of an arbitrary
element of $\frak{s}$, that is $\rho=\rho(s_i)$ with $s_i$
arbitrary. So when we write $\rho(r_1)$ we really mean
$\rho(s_i)(r_1)$. In components we write
\begin{equation}
\rho(r^i)=\rho^i_j r^j\,.
\end{equation}
In what follows we investigate the conditions on the matrix elements
$\rho^i_j$ coming from \eqref{condder}.

\begin{itemize}
\item $\frak{r}=\frak{u}(1)^3$: There are no conditions coming
from (\ref{condder}). \item $\frak{r}=$Heis$_3$:
$\rho^3_2=\rho^3_1=0\,,\quad \rho^3_3=\rho^2_2+\rho^1_1$. Hence only
the $\textbf{3}$ representation is excluded. For the $\textbf{1}
\oplus \textbf{2}$ representation, $r_3$ is the singlet.

\item $\frak{r}=\frak{iso}(2)\, \&\, \frak{iso}(1,1)$:
$\rho^3_1=\rho^2_1=\rho^1_1=0\,, \quad \rho^3_3=\rho^2_2\,, \quad
\rho^2_3=\mp \rho^3_2$, where the upper (lower) sign is for
$\frak{iso}(2)$ ($\frak{iso}(1,1)$). Both the $\textbf{1} \oplus
\textbf{2}$ and the $\textbf{3}$ representation are excluded.

\end{itemize}

Let us summarize  by listing all unipotent algebras that we have
found that are not solvable. First we have the four semi-simple
cases, which we list in table \ref{table:simple}.
\begin{table}[ht]
\centering 
\begin{tabular}{c } 
\hline\hline 
Case  \\ [0.5ex] 
\hline 
$\frak{so}(3)\times \frak{so}(3)$  \\ 
$\frak{so}(3)\times \frak{so}(2,1)$  \\
$\frak{so}(2,1)\times \frak{so}(2,1)$ \\
$\frak{so}(3,1)$ \\ [1ex] 
\hline 
\end{tabular}
\caption{The semi-simple six-dimensional Lie algebras} 
\label{table:simple} 
\end{table}

Then we have the non-semi-simple, non-solvable unipotent examples,
which we list in table \ref{table:nonsimple}.
\begin{table}[ht!]
\centering 
\begin{tabular}{c c } 
\hline\hline 
Case & Representations \\ [0.5ex] 
\hline 
$\frak{so}(3)\ltimes_{\rho} \frak{u}(1)^3$ & $\rho=\textbf{1}
\oplus \textbf{1} \oplus \textbf{1}$ and $\rho=\textbf{3}$  \\ 
$\frak{so}(3)\ltimes_{\rho}$Heis$_3$ &  $\rho=\textbf{1} \oplus
\textbf{1} \oplus \textbf{1}$  \\
$\frak{so}(3)\ltimes_{\rho} \frak{iso}(2)$ & $\rho=\textbf{1}
\oplus \textbf{1} \oplus \textbf{1}$  \\
$\frak{so}(3)\ltimes_{\rho} \frak{iso}(1,1)$ & $\rho=\textbf{1}
\oplus
\textbf{1} \oplus \textbf{1}$  \\
$\frak{so}(2,1)\ltimes_{\rho} \frak{u}(1)^3$ & $\rho=\textbf{1}
\oplus \textbf{1} \oplus \textbf{1}$, $\rho=\textbf{1} \oplus
\textbf{2}$ and $\rho=\textbf{3}$  \\
$\frak{so}(2,1)\ltimes_{\rho}$ Heis$_3$ & $\rho=\textbf{1} \oplus
\textbf{1} \oplus \textbf{1}$ and $\rho=\textbf{1} \oplus
\textbf{2}$  \\
$\frak{so}(2,1)\ltimes_{\rho}\frak{iso}(2)$ & $\rho=\textbf{1}
\oplus \textbf{1} \oplus \textbf{1}$  \\
$\frak{so}(2,1)\ltimes_{\rho}\frak{iso}(1,1)$ & $\rho=\textbf{1}
\oplus \textbf{1} \oplus \textbf{1}$  \\[1ex]  
\hline 
\end{tabular}
\caption{The unipotent non-solvable, non-semi-simple six-dimensional Lie algebras} 
\label{table:nonsimple} 
\end{table}

Hence there are 16 unipotent non-solvable six-dimensional Lie
algebras. These have to be added to the list of solvable unipotent
Lie algebras in \cite{Andriot:2010ju,Grana:2006kf}. As we mentioned
before, unipotence is just one condition for a compactification $L$
to exist when the group $G$ is non-compact, but we do not know which
of these examples cannot be made compact. The non-compact
semi-simple cases of table \ref{table:simple} could be problematic.
When the representation is trivial, $\rho=\textbf{1} \oplus
\textbf{1} \oplus \textbf{1}$, for the cases in table
\ref{table:nonsimple} and the simple part is $\frak{so}(3)$ then we
know for sure that the space can be made compact since $\SO(3)$ is
compact and the three-dimensional solvable groups can be made
compact, see e.g.~\cite[version 1]{Haque:2008jz}.

\section{Discrete symmetries, orbifolds and orientifolds}
\label{sec:discrete}

\subsection{Discrete subgroups of $\SU(3)$}

In this paper we restrict ourselves to the study of
compactifications that preserve $\mathcal{N} =1$ in four dimensions
since these are interesting for phenomenological reasons. Since
group manifolds have a trivial structure group they lead after an
orientifold projection to 4D theories with $\mathcal{N}=4$.
Therefore we have to mod out the group manifolds by a discrete
subgroup of $\SU(3)$ to obtain an $\mathcal{N}=1$ theory. Any dS
critical point we find is then also a dS critical point of the
parent $\mathcal{N}=4$ theory. General studies of 4D $\mathcal{N}=4$
theories \cite{Kallosh:2001gr, deRoo:2002jf, deRoo:2003rm,
Dall'Agata:2009gv, Roest:2009tt, Dibitetto:2010rg, Dibitetto:2011gm}
indicate that they do not allow for a metastable dS solution.
However, when we truncate to an $\mathcal{N}=1$ theory we can hope
that we project out the tachyonic directions and find stable dS
vacua. An unstable dS extremum in the parent $\mathcal{N}=4$ theory
could then be stable in the truncated $\mathcal{N}=1$
theory\footnote{Since we restrict ourselves to $\mathcal{N}=1$
theories we cannot exclude the possibility that one of our group
manifolds leads to (un)stable dS solutions in $\mathcal{N}=2,4$
SUGRA that are not consistent with our orbifold projection i.e.\
that we do not find. It would therefore be very interesting to study
compactifications on group spaces that lead to $\mathcal{N}=2,4$
theories.}. On general grounds one might expect that lowering the
amount of supersymmetry of the effective action also lowers the
constraints on obtaining stable dS critical points. This has been
explicitly confirmed in \cite{Fre:2002pd} by the construction of
stable dS solutions in $\mathcal{N}=2$ gauged supergravity. It is
desirable to understand the higher-dimensional origin of this
solution, as it is unknown as yet.

Let us now explain how we proceed to construct an orientifold to get
an $\mathcal{N}=1$ supergravity. We can define three complex
coordinates $z^i$ on the group spaces. In order to find an
$\mathcal{N}=1$ theory we have to first act on these three complex
coordinates (holomorphically) with a discrete subgroup $\Gamma
\subset$ $\SU(3)$ ($\Gamma \nsubseteq \SU(2)$). Through its action
on the $z^i$, $\Gamma$ also acts on the structure constants and it
is severely constrained by the requirement that these are invariant.
In the case that we have to mod out the group manifold by a lattice
$L$ to make it compact, we also have to ensure that $\Gamma$ maps
the lattice points (i.e.\ points in $G$ that are identified) into
each other. This generically puts strong constraints on the possible
$\Gamma$ for any compact $G/L$. In addition to this orbifold
projection we do an orientifold projection that leads to O6-planes.
The resulting space is the quotient $G/\Gamma/\hat{\sigma}$ with
$\hat{\sigma} = \sigma \Omega_p (-1)^{F_L}$, where $\Omega_p$ is the
world-sheet parity operator and $F_L$ is the space-time fermion
number in the left-moving sector. Furthermore, we have to demand
that $\hat{\sigma}$ is a $\mathbb{Z}_2$ extension of $\Gamma$ i.e.\
that for all $\gamma \in \Gamma$ we have $\gamma \hat{\sigma} \gamma
\hat{\sigma} \in \Gamma$. In order to preserve supersymmetry
$\sigma$ has to act anti-holomorphically on the $z^i$. The 3-cycles
that are fixed under the action of $\sigma$ are wrapped by
O6-planes.

While it is beyond the scope of this paper to list all possible
orbifold and orientifold projections for all group spaces, we will
briefly discuss all discrete subgroups of $\SU(3)$
\cite{discreteSU3, Hanany:1998sd,Ludl:2011gn} that are not contained
in $\SU(2)$. There are two infinite series of abelian subgroups. The
$\mathbb{Z}_N$ subgroups are generated by the following action
\be
\theta: (z^1,z^2,z^3) \rightarrow (\e^{2 \pi i n_1/N} z^1,\e^{2 \pi i
n_2/N}z^2, \e^{2 \pi i (N-n_1-n_2)/N}z^3), \quad n_1 \leq n_2 \leq
N-n_1-n_2.
\ee
One can also combine two cyclic subgroups of $\SU(2)$
to get the subgroup $\mathbb{Z}_N \times \mathbb{Z}_M$ which is
generated by
\subeq{\ba\label{eq:ZnZM}
\theta_1 : (z^1,z^2,z^3) &\rightarrow& (\e^{2 \pi i/N} z^1, \e^{-2 \pi i/N}z^2, z^3),\\
\theta_2 : (z^1,z^2,z^3) &\rightarrow& (\e^{2 \pi i/M} z^1,z^2, \e^{-2 \pi i/M}z^3).
\ea}
There are also two infinite series of non-abelian discrete subgroups
called $\Delta(3n^2)$ and $\Delta(6n^2)$. Both of these groups have
an abelian $\mathbb{Z}_n \times \mathbb{Z}_n$ subgroup.
$\Delta(3n^2)$ is an $\mathbb{Z}_3$ extension of this abelian group
where the $\mathbb{Z}_3$ is generated by the following action \be
(z^1,z^2,z^3) \rightarrow (z^2,z^3,z^1). \ee $\Delta(6n^2)$ is an
$S_3$ extension of $\mathbb{Z}_n \times \mathbb{Z}_n$, where $S_3$
is the permutation group of three elements that is of order 6 and
acts on the three $z^i$. Finally there are the so called exceptional
discrete subgroups of $\SU(3)$: $\Sigma_{36\cdot \varphi},
\Sigma_{60\cdot \varphi},  \Sigma_{72\cdot \varphi},
\Sigma_{168\cdot \varphi}, \Sigma_{216\cdot \varphi},
\Sigma_{360\cdot \varphi}$ with $\varphi=1,3$, where the subscript
for $\varphi=1$ is the order of the group.

If $\Gamma$ has fixed points then $G/\Gamma$ has orbifold
singularities at these fixed points\footnote{These orbifold fixed
points should not be confused with the fixed loci of the orientifold
action. The cycles that are fixed under the elements $\gamma \sigma$
with $\gamma \in \Gamma$ are not singularities but rather the cycles
wrapped by the orientifold sources.}. This is
sensible in string theory where a set of new degrees of freedom,
called the twisted sector, arises at each orbifold singularity. In
order to trust the supergravity analysis one has to resolve the
singularity by ``blowing it up''. This leads to new cycles at the
resolved orbifold singularity. These new cycles and their dual forms
lead to new moduli but also allow us to turn on additional fluxes.
We will not investigate these extra moduli and fluxes and expect
that the extra moduli can be stabilized by the extra fluxes at a
higher scale so that their impact on our analysis is small. This was
explicitly confirmed for a concrete $CY_3$ compactification in
\cite{DeWolfe:2005uu}.

We will now illustrate the general discussion above for the explicit
example of the $\mathbb{Z}_2\times \mathbb{Z}_2$ orbifold since it
plays a prominent role in the construction of dS vacua.

\subsection{The standard $\mathbb{Z}_2\times \mathbb{Z}_2$ orientifold example}\label{sec:standarddetails}

\subsubsection*{The classification of compatible algebras}
In this subsection we work out some details of the
$\mathbb{Z}_2\times \mathbb{Z}_2$ orbifold since this case will be
important in the next section. Rather than taking each group space
$G$ and check whether its structure constants are compatible with
the $\mathbb{Z}_2\times \mathbb{Z}_2$ orbifold action, we work with
a generic group space with one-forms $e^a, a=1,\ldots,6$ that
satisfy $\d e^a = -\tfrac12 f^a{}_{bc} e^b \w e^c$. We check which
structure constants $f^a{}_{bc}$, and therefore which groups, are
compatible with the $\mathbb{Z}_2\times \mathbb{Z}_2$ orbifold
action and a further orientifold action. For non-compact $G$ one
also has to verify the existence and compatibility of a lattice $L$
such that $G/L$ is compact.

Defining the three complex one-forms\footnote{We refrain from using
a more natural definition like $\d z^i = e^i+ \tau_i
e^{i+3},\,i=1,2,3,$ in order to match with \cite{Danielsson:2010bc}
where the authors partially classified group spaces that are
compatible with the $\mathbb{Z}_2\times \mathbb{Z}_2$ orbifolds.}
\ba
\d z^1 &=& e^1 + \tau_1 e^6,\nn \\
\d z^2 &=& e^2 + \tau_2 e^4, \\
\d z^3 &=& e^3 + \tau_3 e^5, \nn \ea where the $\tau_i$ are {\em
complex} complex structure moduli, we find from \eqref{eq:ZnZM} that
the $\mathbb{Z}_2\times \mathbb{Z}_2$ action is generated by
$\theta_1$ and $\theta_2$ that act as \be \theta_1: \left\{
\begin{array}{ccc}
            e^1 & \rightarrow & -e^1 \\
            e^2 & \rightarrow & -e^2 \\
            e^3 & \rightarrow & e^3 \\
            e^4 & \rightarrow & -e^4 \\
            e^5 & \rightarrow & e^5 \\
            e^6 & \rightarrow & -e^6
          \end{array}\right.,\qquad
\theta_2: \left\{ \begin{array}{ccc}
            e^1 & \rightarrow & -e^1 \\
            e^2 & \rightarrow & e^2 \\
            e^3 & \rightarrow & -e^3 \\
            e^4 & \rightarrow & e^4 \\
            e^5 & \rightarrow & -e^5 \\
            e^6 & \rightarrow & -e^6
          \end{array}\right. .
\ee
For this orbifold there are two possible orientifold projections
\cite{Flauger:2008ad}. Here we will focus on the so called {\em
standard orientifold projection} $\sigma: z^i \rightarrow \bar{z}^i,
\, i=1,2,3$. We will present the relevant results for a classification for the
{\em non-standard orientifold projection} $\sigma_{ns}: (z^1,z^2,z^3)
\rightarrow (\bar{z}^1,\bar{z}^3, \bar{z}^2)$ in the next subsection.

An orientifold action always projects out half of the complex
structure moduli. For the standard orientifold projection this
simply results in Re$(\tau_i)=0,\, \forall i$. The explicit action
of $\sigma$ on the $e^i$ is given by \be \sigma: \left\{
\begin{array}{ccc}
            e^1 & \rightarrow & e^1 \\
            e^2 & \rightarrow & e^2 \\
            e^3 & \rightarrow & e^3 \\
            e^4 & \rightarrow & -e^4 \\
            e^5 & \rightarrow & -e^5 \\
            e^6 & \rightarrow & -e^6
            \end{array}\right. .
\ee
This restricts the Lie algebra to be of the particular form
\eq{\spl{
& \d e^1 =f^1{}_{23}e^{23}+f^1{}_{45}e^{45}\,,\qquad \d e^2 =f^2{}_{13}e^{13}+f^2{}_{56}e^{56}\,,\\
& \d e^3 =f^3{}_{12}e^{12}+f^3{}_{46}e^{46}\,,\qquad \d e^4 =f^4{}_{36}e^{36}+f^4{}_{15}e^{15}\,,\label{algebra}\\
& \d e^5 =f^5{}_{14}e^{14}+f^5{}_{26}e^{26}\,,\qquad \d e^6
=f^6{}_{34}e^{34}+f^6{}_{25}e^{25}\,. }} We find that the algebra is
automatically unipotent.

From the combination of the orientifold with the orbifold elements
we find four intersecting O6 planes in the covering space $G$
\begin{center}
  \begin{tabular}{|c|c|c|c|c|c|}
    \hline
    \rule[1em]{0pt}{0pt} $e^1$ & $e^2$  & $e^3$  & $e^4$ & $e^5$  & $e^6$  \\
    \hline
    \hline
    \rule[1em]{0pt}{0pt} $\bigotimes$ & $\bigotimes$  & $\bigotimes$ & -- & -- & -- \\\hline
    \rule[1em]{0pt}{0pt} -- & -- & $\bigotimes$ & $\bigotimes$ & -- & $\bigotimes$ \\\hline
    \rule[1em]{0pt}{0pt} -- & $\bigotimes$ & -- & -- & $\bigotimes$  & $\bigotimes$ \\\hline
    \rule[1em]{0pt}{0pt} $\bigotimes$ & -- & -- & $\bigotimes$  & $\bigotimes$ & -- \\\hline
   \end{tabular}
\end{center}
where each entry denotes a left-invariant direction of $G$.

We want to determine all the unipotent algebras in our
classification that are consistent with the
$\mathbb{Z}_2\times\mathbb{Z}_2$ orbifold and the standard
orientifold.  Let us first do this on the level of the structure
constants and later link it to the algebras in our classification
tables. From now on we group the structure constants into a matrix
denoted $r$ as in \cite{Flauger:2008ad}
\begin{equation}
r=\begin{pmatrix} f^1{}_{45} & f^1{}_{23} & -f^6{}_{34} &-f^6{}_{25}\\
-f^2{}_{56}& f^4{}_{36} & -f^2{}_{13} & -f^4{}_{15}\\
-f^3{}_{46}& f^5{}_{26} & f^5{}_{14} & f^3{}_{12} \end{pmatrix}\,.
\end{equation}

To classify all possible Lie algebras that are invariant under
$\theta_1$, $\theta_2$ and $\sigma$ we need to understand what kind
of $\GL(6,\Real)$-transformations commute with $\theta_1$,
$\theta_2$ and $\sigma$. Then we have to mod out by these
transformations in order to classify inequivalent models. Hence,
consider a general transformation 6 by 6 matrix $M\in\GL(6,\Real)$
that acts on the generators
\begin{equation}
e^i\rightarrow e'^i= M^i_j \,e^j\,.
\end{equation}
The condition that $M$ commutes with $\theta_1$, $\theta_2$ and
$\sigma$ and thus preserves the form of the Maurer--Cartan equations
leads to
\begin{equation}\label{M}
M=\text{diag}(\lambda_1,\lambda_2,\lambda_3,\lambda_4,\lambda_5,\lambda_6)
\end{equation}
where $\lambda_i\neq 0$ corresponds to a rescaling of a generator.
Since there are six $\lambda$'s to choose from one can set up to six
non-zero $f^a{}_{bc}$'s equal to $\pm1$. Then one can check that the Jacobi
identities ensure that the other 6 remaining $f^a{}_{bc}$'s have to be either $\pm 1$ or $0$.
So we have found that the only $M$-transformations that preserve the above form of the
Maurer--Cartan equations are rescalings and that these rescalings
ensure that all structure constants can be normalized to $\pm1,0$.

So now we can start investigating all sets of matrices $r$ with
entries $0, +1, -1$ that obey the Jacobi identities. If we exclude
the trivial algebra then there is at least one structure constant
non-zero. After relabeling of the axes we can always take that to be
the top-left element of $r$, $f^1{}_{45}\neq 0$.

If we go through all Jacobi identities, we find four classes of
allowed algebras: \subeq{\label{eq:classification}\al{
class\,\, A :\,\,  & \quad r=\begin{pmatrix} f^1{}_{45} &  f^1{}_{23} &-f^6{}_{34} & -f^6{}_{25} \\
-f^2{}_{56} & \frac{f^1{}_{23}f^2{}_{56}}{f^1{}_{45}} & -\frac{f^2{}_{56}f^6{}_{34}}{f^1{}_{45}} & \frac{f^6{}_{25}f^2{}_{56}}{f^1{}_{45}} \\
-f^3{}_{46}&  \frac{f^1{}_{23}f^3{}_{46}}{f^1{}_{45}} & \frac{f^6{}_{34}f^3{}_{46}}{f^1{}_{45}} & -\frac{f^3{}_{46}f^6{}_{25}}{f^{1}{}_{45}}  \end{pmatrix}\,,\\
class\,\, B :\,\,  & \quad r=\begin{pmatrix}f^1{}_{45} & 0 &0 & -f^6{}_{25} \\
-f^2{}_{56} & 0 & 0 & -f^4{}_{15} \\
0& 0 & 0 & 0 \end{pmatrix}\,,\\
class\,\, C :\,\,  & \quad r=\begin{pmatrix}f^1{}_{45} &  f^1{}_{23} &-f^6{}_{34} & 0 \\
0 & 0 & 0 & 0 \\
-f^3{}_{46}&  \frac{f^1{}_{23}f^3{}_{46}}{f^1{}_{45}} & f^5{}_{14} & 0\end{pmatrix}\,,\\
class\,\, D :\,\,  & \quad r=\begin{pmatrix}f^1{}_{45} &  0 & 0 & 0 \\
0 & 0 & 0 & -f^4{}_{15} \\
0 & 0 & f^5{}_{14} & 0\end{pmatrix}\,.}} We further discuss some of
these four classes and their corresponding group spaces in section
\ref{dSclassification}.

Note that we obtained the four cases through relabeling of the
generators, in other words by considering permutations. Permutations
are not of the form $M$ as given in equation (\ref{M}) since
permutations change the definition of the various involutions.
However, one can show for the standard orientifold, that
permutations of rows and columns in $r$ map one model to an
equivalent one.

\subsubsection*{The $\SU(3)$-structure}

There are no one-forms that are invariant under the
$\mathbb{Z}_2\times \mathbb{Z}_2$ orbifold action. Furthermore, the
only invariant two-forms have all negative parity under the
orientifold action $\sigma$. We choose them to be
\ba\label{eq:2forms}
Y^{(2-)}_1 &=& \tfrac{i}{2 \hat{\tau}_1} \d z^1 \w \d \bar{z}^1 = e^{16}, \nn \\
Y^{(2-)}_2 &=& -\tfrac{i}{2 \hat{\tau}_2} \d z^2 \w \d \bar{z}^2 = -e^{24}, \\
Y^{(2-)}_3 &=& \tfrac{i}{2 \hat{\tau}_3} \d z^3 \w \d \bar{z}^3 =
e^{35}, \nn
\ea
where $\hat{\tau}_i \equiv \text{Im}(\tau_i)$. The even and odd real three-forms are
\ba\label{eq:3forms}
Y^{(3+)}_1 &=& e^{123}, \quad Y^{(3+)}_2 = e^{145}, \quad Y^{(3+)}_3 = -e^{256}, \quad Y^{(3+)}_4 = -e^{346},  \\
Y^{(3-)1} &=& e^{456}, \quad Y^{(3-)2} = e^{236}, \quad Y^{(3-)3} =
e^{134}, \quad Y^{(3-)4} = e^{125}.
\ea
Since $J$ and $\Omega_R$ must be odd under the $\SU(3)$-structure preserving
orientifold involution, we find that they must be of the form\footnote{Note,
that we have relabeled the expansion coefficient and
flipped the sign in front of $e^{24}$ compared to
\cite{Danielsson:2010bc}.} \subeq{\label{JOmexp}
\begin{align}
& J = k^i Y^{(2-)}_i = k^1 e^{16}- k^2 e^{24} + k^3 e^{35}\,,\\
&\Omega_R =\mathcal{F}_K Y^{(3-)K} = \mathcal{F}_1 e^{456} + \mathcal{F}_2e^{236} + \mathcal{F}_3e^{134} + \mathcal{F}_4e^{125}\,,
\end{align}}
where $k^i$ and $\mathcal{F}_K$ are real coefficients.

From
\begin{equation}
J\wedge J\wedge J = 6 \,k^1 k^2 k^3 \, e^{123456}\,,
\end{equation}
we find that (for our choice of orientation) $k^1 k^2 k^3>0$
rendering all or one of the coefficients $k^i$ positive. This allows
us to compute the complex structure $I$. In order to be able to
properly normalize it with real $c$ in \eqref{complexstructure} we need furthermore
$\mathcal{F}_1\mathcal{F}_2\mathcal{F}_3\mathcal{F}_4>0$. Explicitly
we find
\begin{align}
& I^1_6=-\sqrt{\frac{\mathcal{F}_1\mathcal{F}_2}{\mathcal{F}_3\mathcal{F}_4}}\,,\qquad I^6_1=\sqrt{\frac{\mathcal{F}_3\mathcal{F}_4}{\mathcal{F}_1\mathcal{F}_2}}\,, \nonumber \\
& I^2_4=+\sqrt{\frac{\mathcal{F}_1\mathcal{F}_3}{\mathcal{F}_2\mathcal{F}_4}}\,,\qquad I^4_2=-\sqrt{\frac{\mathcal{F}_2\mathcal{F}_4}{\mathcal{F}_1\mathcal{F}_3}}\,,  \\
&
I^3_5=-\sqrt{\frac{\mathcal{F}_1\mathcal{F}_4}{\mathcal{F}_2\mathcal{F}_3}}\,,\qquad
I^5_3=\sqrt{\frac{\mathcal{F}_2\mathcal{F}_3}{\mathcal{F}_1\mathcal{F}_4}}\,.\nonumber
\end{align}
From equation (\ref{metricfromI}) we obtain the metric, which turns
out to be diagonal, consistent with even parity under the
orientifold involution
\begin{equation}
g=\frac{1}{\sqrt{\mathcal{F}_1\mathcal{F}_2\mathcal{F}_3\mathcal{F}_4}}\Bigl(k^1 \mathcal{F}_3\mathcal{F}_4\,,\, k^2 \mathcal{F}_2\mathcal{F}_4\,,\, k^3 \mathcal{F}_2\mathcal{F}_3\,,\, k^2 \mathcal{F}_1\mathcal{F}_3\,,\, k^3 \mathcal{F}_1\mathcal{F}_4\,,\, k^1 \mathcal{F}_1\mathcal{F}_2\Bigr)\,.
\end{equation}
With the metric available we can compute $\Omega_I=\star\Omega_R$
\begin{equation}
\Omega_I=\mathcal{Z}^K Y^{(3+)}_K = \sqrt{\mathcal{F}_1\mathcal{F}_2\mathcal{F}_3\mathcal{F}_4}
\Bigl(\mathcal{F}_1^{-1}\,e^{123}+\mathcal{F}_2^{-1}\,e^{145}-\mathcal{F}_3^{-1}\,e^{256}-\mathcal{F}_4^{-1}e^{346}\Bigl)\,.
\end{equation}
The normalization condition \eqref{normcond} leads to
\begin{equation}
\sqrt{\mathcal{F}_1\mathcal{F}_2\mathcal{F}_3\mathcal{F}_4}=k^1k^2k^3\,.
\end{equation}
The normalization condition \eqref{normcond} also allows us to determine the normalization of
\be
\Omega = \mathcal{F}_K Y^{(3-)K} + i \mathcal{Z}^K Y^{(3+)}_K \propto \, \d z^1 \w \d z^2 \w \d z^3
\ee
so that we can express the $\mathcal{Z}^K$ and $\mathcal{F}_K$ explicitly in terms of the three independent complex structure parameters $\hat{\tau}_i \equiv \text{Im}(\tau_i)$ and the $k^i$
\ba
\mathcal{Z}^1 &=& \frac{1}{\sqrt{\hat{\tau}_1 \hat{\tau}_2 \hat{\tau}_3}}, \quad \mathcal{Z}^2 =- \sqrt{\frac{\hat{\tau}_2\hat{\tau}_3}{\hat{\tau}_1}}, \quad \mathcal{Z}^3 = \sqrt{\frac{\hat{\tau}_1\hat{\tau}_3}{\hat{\tau}_2}}, \quad \mathcal{Z}^4 = -\sqrt{\frac{\hat{\tau}_1\hat{\tau}_2}{\hat{\tau}_3}}, \\
\mathcal{F}_K &=& \frac{\sqrt{k^1 k^2 k^3}}{\mathcal{Z}^K}, \quad \forall K.
\ea
The absence of nowhere-vanishing one-forms on homogeneous strict
$\SU(3)$-structure spaces implies $W_4 = W_5 = 0$ \footnote{For non-homogeneous
SU(3)-structures this is not true as forms $W_4$ and $W_5$ that vanish
somewhere are still allowed.}. Since there are no
even two-forms in this example there are no gauge fields and no
D-terms, so that we find that $W_1,W_2$ are real and we obtain a
half-flat $\SU(3)$-structure space (cf.\ appendix \ref{halfflat}). Furthermore,
we can construct the remaining torsion classes from the
identities\footnote{To understand how the torsion classes depend on
all moduli is not too hard for this simple example, but formulae for
more general cases have been derived in \cite{Ihl:2007ah}.}
\subeq{\al{ W_1&=-\tfrac{1}{6}\star_6 (\d J \wedge\Omega_I)\,,\\
W_2&=-\star \d\Omega_I+2W_1J\,,\\ W_3&=\d
J-\tfrac{3}{2}W_1\Omega_R\,.}}


\subsection{The non-standard  $\mathbb{Z}_2\times \mathbb{Z}_2$ orientifold example} \label{sec:nonstandarddetails}
We provide some details for an alternative orientifold projection of the
$\mathbb{Z}_2\times \mathbb{Z}_2$ orbifold, which leads to explicit
examples of 4D supergravity theories with D-terms.

We take the $\mathbb{Z}_2 \times \mathbb{Z}_2$ orbifold group to act
on the six one-forms $e^a$ as \be \theta_1: \left\{ \begin{array}{ccc}
            e^1 & \rightarrow & -e^1 \\
            e^2 & \rightarrow & -e^2 \\
            e^3 & \rightarrow & e^3 \\
            e^4 & \rightarrow & -e^4 \\
            e^5 & \rightarrow & -e^5 \\
            e^6 & \rightarrow & e^6
          \end{array} \right.,
\quad \qquad \theta_2: \left\{\begin{array}{ccc}
            e^1 & \rightarrow & -e^1 \\
            e^2 & \rightarrow & e^2 \\
            e^3 & \rightarrow & -e^3 \\
            e^4 & \rightarrow & -e^4 \\
            e^5 & \rightarrow & e^5 \\
            e^6 & \rightarrow & -e^6
          \end{array} \right..
\ee Due to the simplicity of this action we see that it does not
relate any of the $f^a{}_{bc}$ to each other, rather it projects out
all but 24 of the metric fluxes. In the case of the standard
orientifold projection 12 more metric fluxes get projected out by
the orientifold projection. However, for the non-standard
orientifold projection $\sigma_{\text{ns}}$ we have the following
action on the $e^a$ \be \sigma_{\text{ns}}: \left\{
\begin{array}{ccc}
            e^1 & \rightarrow & e^1 \\
            e^2 & \rightarrow & e^3 \\
            e^3 & \rightarrow & e^2 \\
            e^4 & \rightarrow & -e^4 \\
            e^5 & \rightarrow & -e^6 \\
            e^6 & \rightarrow & -e^5
          \end{array} \right..
\ee $\sigma_{\text{ns}}$ relates 10 of the metric fluxes to each
other \ba
\begin{array}{lll}
  f^1{}_{26}=- f^1{}_{35}, &\qquad \qquad& f^4{}_{35}= f^4{}_{26}, \\
  f^2{}_{13}= f^3{}_{12}, && f^5{}_{13}=- f^6{}_{12}, \\
  f^2{}_{16}=- f^3{}_{15}, && f^5{}_{16}= f^6{}_{15}, \\
  f^2{}_{34}=- f^3{}_{24}, && f^5{}_{34}= f^6{}_{24}, \\
  f^2{}_{46}= f^3{}_{45}, && f^5{}_{46}=- f^6{}_{45},
\end{array}
\ea and projects out two: $f^1{}_{23} = f^1{}_{56}=0$.

It is straightforward to check that we have two odd and one even
two-form which we take to be \be Y^{(2-)}_1 = -\frac12 e^{14}, \qquad
Y^{(2-)}_2 = e^{25} + e^{36}, \qquad Y^{(2+)} =  e^{25} - e^{36}.
\ee So in this case we have two matrices $r_{iK}$ and $\hat{r}_K$
such that \be \d Y^{(2-)}_i = -r_{iK} Y^{(3-)K},\, i=1,2, \qquad \d
Y^{(2+)} = -\hat{r}^{K} Y^{(3+)}_K, \, K=1,2,3,4.
\ee
Choosing the three-forms to be \ba
\begin{array}{lll}
  Y^{(3+)}_1 = e^{126}-e^{135}, & \qquad \qquad & Y^{(3-)1} = \frac12 \lp e^{246} -e^{345}\rp, \\
  Y^{(3+)}_2 = -e^{234}, & & Y^{(3-)2} = e^{156}, \\
  Y^{(3+)}_3 = -e^{456}, & & Y^{(3-)3} = e^{123}, \\
  Y^{(3+)}_4 = -e^{246}-e^{345}, & & Y^{(3-)4} = -\frac12 \lp e^{126}+e^{135} \rp,
\end{array}
\ea we find the explicit expressions \ba r &=& \left(
     \begin{array}{cccc}
       f^1_{26} & \frac12 f^4_{56} & \frac12 f^4_{23} & -f^4_{26}\\
       2 (f^3_{24} + f^6_{45}) & 2 f^3_{15} & -2 f^6_{12} & -2 (f^3_{12} + f^6_{15})
     \end{array}
   \right), \\
\hat{r} &=& \left(
            \begin{array}{cccc}
              -f^3_{12} + f^6_{15} & 2 f^6_{24} & 2 f^3_{45} & f^3_{24} - f^6_{45} \\
            \end{array}
          \right).
\ea
The matrices are again constrained by the requirement that
$\d^2 e^a \propto f^a{}_{b[c}f^b{}_{de]} = 0,\, \forall a$. This
gives in our case 11 independent constraints; only two of these are
obtained by demanding that $\d^2 Y^{(2-)}_i = r_{iK} \hat{r}^K
Y^{(4+)}=0,\, i=1,2$. Furthermore, we have to satisfy $\d H = h_K \d
Y^{(3-)K} = h_K \hat{r}^K Y^{(4+)}=0$, i.e.\ the metric fluxes have
to satisfy $h_K \hat{r}^K=0$.

\subsection{General abelian and non-abelian orbifolds}

Contrary to for example the torus $\U(1)^6= T^6 =\Real^6/L$, for any compact
group space it is not necessary to mod out by a lattice $L$ so that for these group spaces the only
constraint on the orbifold group is that it is compatible with the
structure constants. This means that in principle all the discrete subgroups of $\SU(3)$ should be studied in the search for dS vacua. In \cite{Flauger:2008ad} the authors investigated the existence of dS extrema in 17 abelian orbifolds of small order and found that all of their examples except $\mathbb{Z}_2\times \mathbb{Z}_2$ can be excluded by no-go theorems. Intuitively the reason for this seems to be that quotients by orbifold groups of large order lead to models with only very few moduli and structure constant. $\mathbb{Z}_2\times \mathbb{Z}_2$ is the most symmetric case which allows for the largest number of moduli and structure constants and might therefore be the only abelian orbifold that leads to dS vacua. However, there is no proof of this and it is also possible that an abelian orbifold group that has not been studied yet allows for dS vacua. We do not pursue this here but leave it to the interested reader to construct such models or find no-go theorems.

Non-abelian orbifolds have not been studied at all in the context of
de Sitter model building.\footnote{The so called isotropic $\mathbb{Z}_2\times \mathbb{Z}_2$ orbifold that was studied in \cite{deCarlos:2009fq,deCarlos:2009qm} is the non-abelian orbifold $\Delta(12)$.} Here, we would like to point out that they
can be excluded if they contain an abelian subgroup that has already been excluded. So for example: based on the results of
\cite{Flauger:2008ad}, we find that $\Delta(3n^2)$ and $\Delta(6n^2)$ are excluded for $n=3,4,6$. We will discuss the interesting case $n=2$ below. Since there are only finitely many forms and structure constants on a six-dimensional group space, one expects, as we argued above, that discrete subgroups of small order are the most promising because they lead to the richest set of fields and structure constants. Based on this observation it seems unlikely that $\Delta(3n^2)$ and $\Delta(6n^2)$ for $n=5,7,8,\ldots$ or any of the exceptional non-abelian groups (which are of order 36 or larger) can lead to dS vacua. Therefore we will focus on $\Delta(12)$ which is important for our construction of dS vacua in the next section.

\subsubsection{The standard orientifold of $\Delta(12)$}
\label{delta12stand}
For the standard orientifold projection, we find that the
$\mathbb{Z}_3$ extension of $\mathbb{Z}_2\times \mathbb{Z}_2$ acts
on the $e^i$ as\footnote{The two extra minus signs arise due to the minus sign in our definition of $Y^{(2-)}_2$ in \eqref{eq:2forms}.}
\be
\mathbb{Z}_3: \left\{ \begin{array}{ccc}
            e^1 & \rightarrow & e^2 \\
            e^2 & \rightarrow & e^3 \\
            e^3 & \rightarrow & e^1 \\
            e^4 & \rightarrow & -e^5 \\
            e^5 & \rightarrow & e^6 \\
            e^6 & \rightarrow & -e^4
            \end{array}\right. .
\ee
This leads to an identification of the moduli in \eqref{JOmexp}
so that there is only one K\"ahler modulus $k\equiv k^1=k^2=k^3$ and two
complex structure parameter $\mathcal{F}_1$ and $\mathcal{F}_2=
\mathcal{F}_3=\mathcal{F}_4$. Furthermore, the structure constants
have to satisfy the following additional constraints \ba
f^1{}_{45} &=&-f^2{}_{56}=-f^3{}_{46},\\
f^1{}_{23} &=&-f^2{}_{13}=f^3{}_{12},\\
f^6{}_{34} &=&f^4{}_{15}=-f^5{}_{26},\\
f^6{}_{25} &=&-f^4{}_{36}=-f^5{}_{14}. \ea
The Jacobi identities can only be  solved if $f^1{}_{23} = f^6{}_{34} =
f^6{}_{25}$ so that we find the $r$-matrix
\be r=\left(
    \begin{array}{cc}
      f^1{}_{45} & -f^1{}_{23} \\
    \end{array}
  \right).
\ee
The structure constants can again be rescaled so that they are $0, \pm1$.

\subsubsection{The non-standard orientifold of $\Delta(12)$}
\label{delta12nonstand}
For the non-standard orientifold projection (see
\ref{sec:nonstandarddetails} for details) the extension
$\mathbb{Z}_3$ acts as
\be
\mathbb{Z}_3: \left\{ \begin{array}{ccc}
            e^1 & \rightarrow & e^2 \\
            e^2 & \rightarrow & e^3 \\
            e^3 & \rightarrow & e^1 \\
            e^4 & \rightarrow & e^5 \\
            e^5 & \rightarrow & e^6 \\
            e^6 & \rightarrow & e^4
            \end{array}\right. .
\ee
This quotient projects out the even two-form so that there is no
gauge field and therefore no D-term anymore. We are again left with
a model that has one K\"ahler modulus and two complex structure
parameters. The extra $\mathbb{Z}_3$ quotient requires that the structure constants satisfy
the constraints
\ba
f^1{}_{26} &=& f^3{}_{15}=-f^3{}_{24},\\
f^4{}_{26} &=& f^6{}_{15}= f^6{}_{24},\\
f^4{}_{23} &=& f^6{}_{12},\\
f^4{}_{56} &=& f^6{}_{45},\\
f^3{}_{12} &=& f^3{}_{45}=0.
\ea
This case is peculiar since there are four independent structure constants but only two linear
combinations appear in the $r$-matrix
\be r=\left(
    \begin{array}{cc}
      2 f^1{}_{26}-f^4{}_{56} & f^4{}_{23} \\
    \end{array}
  \right).
\ee
The existence of extra structure constants that do not appear in the $r$- or $\hat{r}$-matrices, and therefore not in the scalar potential, was previously noticed for some abelian-orbifolds in \cite{Flauger:2008ad}. These extra structure constants appear in the Jacobi identities. For $r\neq 0$ the Jacobi identities can only be solved for
\ba
f^4{}_{26} &=&0, \quad f^1{}_{26}=0, \quad f^4{}_{23}=0, \\
\text{or} \quad f^4{}_{26} &=&0, \quad f^1{}_{26}=f^4{}_{56}.
\ea

\section{Classification of de Sitter solutions}\label{deSitterSolutions}
As we have argued at the end of section \ref{4d}, flux
compactifications of massive type IIA together with O6-planes lead
to a 4D scalar potential that depends on all moduli. Among the
moduli there are two universal ones that are present for any
compactification space, the volume modulus $\rho
=(\text{vol}_6)^{1/3}$ and the dilaton $\phi$. Based on the scaling
behavior of the scalar potential with respect to these two fields,
the authors of \cite{Hertzberg:2007wc} showed that type IIA flux
compactification with RR- and $H$-flux and O6-planes cannot lead to
dS vacua when the compact space is a $CY_3$ manifold. However, this
no-go theorem can be circumvented on $\SU(3)$-structure spaces with
negative scalar curvature. This observation led to many papers
\cite{Silverstein:2007ac, Haque:2008jz, Caviezel:2008ik,
Flauger:2008ad, Caviezel:2009tu, Danielsson:2009ff, Wrase:2010ew}
that generalized the no-go theorem of \cite{Hertzberg:2007wc} but
also to constructions of actual classical dS solutions from the 4D
effective theory \cite{Saltman:2004jh, Silverstein:2007ac,
Haque:2008jz, Flauger:2008ad, Caviezel:2008tf, deCarlos:2009fq,
deCarlos:2009qm, Caviezel:2009tu, Dong:2010pm, Andriot:2010ju} (see
also \cite{Dibitetto:2010rg}) as well as 10D constructions
\cite{Danielsson:2009ff, Danielsson:2010bc}. However, so far all
explicit constructions of geometric dS solutions are perturbatively
unstable since there is at least one tachyonic field. In
compactifications on more general non-geometric spaces it is
possible to find stable dS vacua \cite{deCarlos:2009fq,
deCarlos:2009qm}. However, since the internal space is non-geometric
it is not clear whether the supergravity approach is reliable.

In \cite{Flauger:2008ad} the authors studied twisted versions of all
abelian toroidal orbifolds $T^6/\mathbb{Z}_N$ and $T^6/\mathbb{Z}_N
\times \mathbb{Z}_M$ and checked whether they allow for dS solutions
or can be excluded by no-go theorems. Since the orbifold and
orientifold actions have to be compatible with the torus lattice, it
turns out that there are 36 different cases
\cite{Reffert:2006du,Reffert:2007im,Flauger:2008ad}. However, if one
restricts to the untwisted sector\footnote{It would be very
interesting to study the inclusion of the twisted sector. This has
not been done in the literature and we refrain from doing it since
the number of moduli would increase substantially, thus making the
analysis of the scalar potential very complicated.} only 11 of these
models are different. Out of these 11 models only the two different
orientifolds of the $\mathbb{Z}_2 \times \mathbb{Z}_2$ orbifold
could not be excluded by no-go theorems and were shown to allow for
dS solutions. Only for the non-standard orientifold projection were
the authors able to construct an explicit compact example. For the
standard orientifold projection it was shown in
\cite{Caviezel:2008tf} that the group space $\SU(2)\times \SU(2)$
provides an explicit compact example that allows for dS solutions.
In this section we will fully classify all possible group spaces
that lead to dS solutions and that are compatible with the
$\mathbb{Z}_2 \times \mathbb{Z}_2$ orbifold action and either one of
the orientifold projections. As in the previous papers we establish
the solutions numerically by minimizing the slow roll parameter
\begin{equation}
\epsilon = G^{IJ}\frac{\partial_I V\partial_J V}{2V^2}\,,
\end{equation}
where $G^{IJ}$ denotes the inverse field space metric.

\subsection{The standard $\mathbb{Z}_2\times \mathbb{Z}_2$ orientifold}\label{dSclassification}
In the previous section in equation \eqref{eq:classification} we
have classified the group manifolds that are compatible with the
$\mathbb{Z}_2\times \mathbb{Z}_2$ orbifold and the standard
orientifold projection. In this section we check which of these
group spaces can lead to dS vacua\footnote{The Mathematica package
STRINGVACUA \cite{Gray:2008zs} is very useful in calculating the
scalar potential for these models.}.

Class B and C are excluded by a no-go theorem \cite{Flauger:2008ad}
since they have at least one zero row. Class D is of the form
$G_6=G_3\times U(1)^3$, with $G_3$ a unipotent three-dimensional
algebra. For this case there is no analytic no-go theorem, but
numerically one finds $\epsilon \approx 1.57221$ unless
$f^1{}_{45}=f^4{}_{15}=f^5{}_{14}$ in which case one gets $\epsilon
\approx 2$ (cf.\ \cite{Flauger:2008ad}). For class A we make further
subdivisions: The no-go theorem implies that $f^2{}_{56}$ and
$f^3{}_{46}$ should be non-zero, otherwise the second, resp. the
third row would vanish. Hence we have the following rule:
\begin{equation}
f^2{}_{56}, \, f^3{}_{46} \in \{-1, +1\}\,,\qquad f^1{}_{23}, \, f^6{}_{34}, \,
f^6{}_{25} \in \{-1, 0, +1\}\,.
\end{equation}
When any of the ($f^1{}_{23}, f^6{}_{34}, f^6{}_{25}$) is zero, then
the whole column is zero. Hence we consider the following subclasses
\begin{itemize}
\item class AI  : all entries non-zero.
\item class AII : one column zero.
\item class AIII: two columns zero.
\item class AIV : three columns zero.
\end{itemize}
For class AIV there is no no-go theorem but numerically one finds $\epsilon \approx \frac{4}{3}$ (cf.\ \cite{Flauger:2008ad}).

Class AI comprise the three simple algebras: $\frak{so}(4)$,
$\frak{so}(2,2)$ and $\frak{so}(3,1)$, as can be easily verified by
computing the Cartan--Killing metric. For these three algebras we
can numerically get $\epsilon$ to vanish. Explicitly we have
\eq{\spl{ r_{\frak{so}(4)} &=
\begin{pmatrix}
+1 & +1 & +1 & -1 \\
+1 & -1 & -1 & -1 \\
+1 & -1 & +1 & +1
\end{pmatrix}\,,\qquad
r_{\frak{so}(2,2)} = \begin{pmatrix}
+1 & +1 & +1 & -1 \\
-1 & +1 & +1 & +1 \\
+1 & -1 & +1 & +1
\end{pmatrix}\,,\\
r_{\frak{so}(3,1)} &=\begin{pmatrix}
-1 & +1 &+1 & -1 \\
-1 & -1 &-1 & -1 \\
-1&  -1 &+1 & +1
\end{pmatrix}\,,}}
up to re-scalings of generators with $\pm 1$.

Class AII consists out of algebras that are all not solvable,
neither semi-simple, they are in between. It is sufficient to take
the last column zero and the first three-non-zero. For instance,
assume it is the third column instead that is zero. Then, by the
permutation $(e_3, e_4) \leftrightarrow (e_2, e_5)$ we can push all
zeros to the fourth column. Furthermore, by fixing the scaling of
the generators $e_1, e_2, e_3$ we can always take the first column
to consist of $+1$'s. A remaining scaling freedom is a
simultaneous rescaling of $e_4, e_5, e_6$ with $\pm 1$. So we find
the algebras
\begin{align}
r_{a} &=\begin{pmatrix}
+1 & +1 &+1 & 0 \\
+1 & -1 &-1 & 0 \\
+1 & -1 &+1 & 0
\end{pmatrix}\,,\qquad
r_{b} =\begin{pmatrix}
+1 & +1 &-1 & 0 \\
+1 & -1 &+1 & 0 \\
+1 & -1 &-1 & 0
\end{pmatrix}\,,\\
r_{c} &=\begin{pmatrix}
+1 & -1 &+1 & 0 \\
+1 & +1 &-1 & 0 \\
+1 & +1 &+1 & 0
\end{pmatrix}\,,\qquad
r_{d} =\begin{pmatrix}
+1 & -1 &-1 & 0 \\
+1 & +1 &+1 & 0 \\
+1 & +1 &-1 & 0
\end{pmatrix}\,.
\end{align}
All of these give vanishing $\epsilon$. These algebras  must be
contained in table \ref{table:nonsimple}. One can verify that
generators $T_3, T_4, T_6$ form the semi-simple piece and that the
generators $T_1, T_2, T_5$ form the radical. It is then
straightforward to read off the representation $\rho$. We find that
\begin{align}
&\frak{so}(3)\ltimes_3 \frak{u}(1)^3 \quad\text{is given by}\quad r_a\,,\nonumber\\
&\frak{so}(2,1)\ltimes_3 \frak{u}(1)^3 \quad \text{is given by}\quad
r_b, r_c, r_d\,.
\end{align}
Thus, only two out of these four sets of structure constants are independent.

Class AIII are all solvable algebras. We can take, without loss of
generality, the last two columns to be zero and take the first
column to be $+1$. Then, we find the following two algebras
\begin{equation}
r_{solvable1} =\begin{pmatrix}
+1 & +1 & 0 & 0 \\
+1 & -1 & 0 & 0 \\
+1 & -1 & 0 & 0
\end{pmatrix}\,,\qquad
r_{solvable2} =\begin{pmatrix}
+1 & -1 & 0 & 0 \\
+1 & +1 & 0 & 0 \\
+1 & +1 & 0 & 0
\end{pmatrix}\,.
\end{equation}
Both of these give vanishing $\epsilon$. Solvable2 equals the
algebra s1.2 as can be seen by rescaling $e_1$ and $e_2$ by $-1$ to
obtain the expression of \cite{Danielsson:2010bc}. This case is the
dS example with the standard orientifold of \cite{Flauger:2008ad}.
The second solvable algebra does not appear in the table of
\cite{Grana:2006kf} and implies that it is not algebraic. But, as
indicated in \cite{Grana:2006kf} also s1.2 is not known to be
algebraic. As a consequence we do not know whether these
solvmanifolds can be compactified by dividing by an appropriate
lattice $L$.

\subsection{The non-standard $\mathbb{Z}_2\times \mathbb{Z}_2$ orientifold}

This case is substantially less symmetric than the standard
orientifold projection so that we refrain from explicitly spelling out every
single solution to the Jacobi identities in terms of the $r$ and
$\hat{r}$ matrices. However, it is straightforward and analogous to the previous
subsection to work out all the details. Two of the Jacobi
identities are \be r_{11} \hat{r}^1 =0,\qquad r_{11} r_{21} =0. \ee
This suggest the classification:
\begin{itemize}
\item $A_{ns}: r_{11} \neq 0, \, r_{21} =0, \, \hat{r}^1 =0$
\item $B_{ns}: r_{11} = 0, \, r_{21} =0, \, \hat{r}^1 \neq 0$
\item $C_{ns}: r_{11} = 0, \, r_{21} \neq 0, \, \hat{r}^1 =0$
\item $D_{ns}: r_{11} = 0, \, r_{21} \neq 0, \, \hat{r}^1 \neq0$
\item $E_{ns}: r_{11} = 0, \, r_{21} =0, \, \hat{r}^1 =0$
\end{itemize}
For each of these classes we have to satisfy an additional 9 Jacobi
identities and the constraint that $h_K \hat{r}^K=0$. It is
straightforward to work out the generic solution for each class.
Since we are interested in the existence of dS vacua we recall the
no-go theorems derived in \cite{Flauger:2008ad} which tell us that
dS extrema and slow-roll inflation are excluded if one of these
three conditions is satisfied
\begin{itemize}
\item $r_{1K}=0, \forall K$,
\item $r_{2K} = \hat{r}^K = 0, \forall K$,
\item $r_{a1} = h_1 = \hat{r}^L =0$ for $a=1,2$, $L=2,3,4$.
\end{itemize}
For the class $B_{ns}$ the 9 additional Jacobi identities require $\hat{r}^2 = \hat{r}^3 =\hat{r}^4=0$ and $h_1=0$ so that this entire class is excluded.
In all of the other classes there are always solutions that fall under one of the above no-go theorems so that these cases do not lead to dS vacua. However, for classes $A_{ns}$, $D_{ns}$, $E_{ns}$ we were able to find dozens
of numerical dS solutions that correspond to a variety of group spaces.

In order to recognize equivalent models, one should divide out by
those transformations that commute with the orbifold and orientifold
involutions. Explicitly these transformations are given by a
six-parameter family of matrices (such that det$M \neq 0$)
\begin{equation}
M=\begin{pmatrix}
\lambda_1 & 0        & 0          & 0        & 0         &0\\
0         &\lambda_2 & 0          & 0        & \lambda_3 & 0\\
0         & 0        & \lambda_2  & 0        & 0         & -\lambda_3\\
0         & 0        & 0          & \lambda_4& 0         & 0 \\
0         & \lambda_5& 0          & 0        & \lambda_6 & 0\\
0         & 0        & -\lambda_5 & 0        & 0         & \lambda_6
\end{pmatrix}\,.
\end{equation}
In the standard case there were also transformations that did not
commute with the involutions but that still mapped equivalent models
to each other. This does not happen for the non-standard orientifold
projection and complicates the analysis.

We find that the semi-simple algebras $\frak{so}(4), \frak{so}(3,1)$ and $\frak{so}(2,2)$ again allow for
numerical de Sitter solutions. Of the non-solvable, non semi-simple
algebras we found de Sitter solutions on $\frak{so}(2,1)\times
\frak{iso}(1,1)$. Apart from that we have numerically found several examples of
solvable algebras that allow de Sitter solutions, all are solvable
of order 3 except one which is solvable of order 2. A typical example
that is solvable of order 3 is
\begin{align}
& f^1_{26}=1\,,\quad f^1_{35}=-1\,,\quad
\quad f^{2}_{46}=-\tfrac{1}{2}\,,\nonumber\\
&  f^3_{45}=-\tfrac{1}{2}\,,\quad f^5_{34}=\tfrac{1}{2}\,,\quad
f^{6}_{24}=\tfrac{1}{2}\,.
\end{align}
The only example that is solvable of order 2 reads
\begin{align}
& f^1_{26}=1\,,\quad f^1_{35}=-1\,,\quad
f^2_{16}=-\tfrac{1}{2}\,,\quad f^{2}_{46}=1\,,\nonumber\\
& f^3_{15}= \tfrac{1}{2}\,,\quad f^3_{45}=1\,,\quad
f^4_{26}=-1\,,\quad f^{4}_{35}=-1\,.
\end{align}
This example was already presented in \cite{Flauger:2008ad}. Since
it was constructed using the so-called base-fiber method it is
manifestly compact.

Again, none of the nilpotent algebras allow de Sitter solutions.
This was the same for the standard orientifold projection and one
could wonder whether there is some no-go theorem that forbids de Sitter solutions
for these spaces. The fact that nilmanifolds seem excluded is rather
surprising since these manifolds are negatively curved for all
values of the moduli, as opposed to the solvable cases or the
semi-simple cases.

\subsection{The non-abelian orbifold $\Delta(12)$}
In order to construct dS solutions that are as simple as possible,
one wants to investigate the possibility of taking a further
quotient of the above models. This potentially allows for dS
solutions with a smaller number of fields that are easier to
analyze. As we discussed above $\Delta(3n^2)$ and $\Delta(6n^2)$ for
$n=2$ are extensions of the abelian group $\mathbb{Z}_2\times
\mathbb{Z}_2$. It is straightforward to check that there are no
three-forms that are compatible with $\Delta(24)$ and either of
our two different orientifolds so that we will focus on $\Delta(12)$.

\subsubsection*{The standard $\Delta(12)$ orientifold}
There are four different cases (see subsection \ref{delta12stand} for details):
For $f^1{}_{45} =f^1{}_{23} =\pm1$ the algebra is $\frak{so}(4)$, for $f^1{}_{45}=
-f^1{}_{23} = \pm1$ the algebra is $\frak{so}(3,1)$, for
$f^1{}_{45}=0$, $f^1{}_{23}\neq 0$ we have $\frak{so}(3)\ltimes_3
\frak{u}(1)^3$ and finally for $f^1{}_{45} \neq0$, $f^1{}_{23}= 0$
we have case AIV from above. Numerically we find $\epsilon \geq
\tfrac{27}{13}$ for $\frak{so}(3)\ltimes_3 \frak{u}(1)^3$ and $\epsilon \geq \tfrac43$
for $\frak{so}(3,1)$ as well as the case AIV
but we are able to find numerical dS solutions for $\frak{so}(4)$.
The latter could have been anticipated since the dS solutions of
\cite{Caviezel:2008tf, deCarlos:2009fq, deCarlos:2009qm,
Danielsson:2010bc} are found by setting all K\"ahler and three
complex structure moduli equal so that they are not only dS
solutions of $\SU(2)\times\SU(2)/\mathbb{Z}_2\times \mathbb{Z}_2$ but also of
$\SU(2)\times\SU(2)/\Delta(12)$.

\subsubsection*{The non-standard $\Delta(12)$ orientifold}
In this case the existence of dS extrema depends only on the values of $2 f^1{}_{26}-f^4{}_{56}$ and
$f^4{}_{23}$ and not on $2 f^1{}_{26}+f^4{}_{56}$ and $f^4{}_{26}$ as discussed in subsection \ref{delta12nonstand}.
Numerically (up to rescalings) we only find dS extrema for $2 f^1{}_{26}-f^4{}_{56}= f^4{}_{23} = \pm1$. This was expected since the scalar potential for the standard and non-standard $\Delta(12)$ orientifold are identical and in the standard case there was only one choice for the $r$-matrix that allowed for dS solutions. The requirement that $f^4{}_{23} \neq 0$ together with the Jacobi identities (cf.\ \ref{delta12nonstand}) fully determines the algebra to be $\frak{so}(2,2)=\frak{so}(2,1) \times \frak{so}(2,1)$.

\subsection{The mass spectrum}
The explicit models we have discussed have up to 14 real moduli. In
order for our dS critical points to be stable we have to demand that
the masses of all these moduli are positive\footnote{Since we
restrict to left-invariant forms, it is possible that there are
unstable modes that are not captured by our consistent truncation.}.
Since there is no supersymmetry that could ensure stability, the
na\"ive expectation is that roughly half of the fields are tachyonic
for a generic critical point. However, this is not the case. It
turns out that all numerical solutions we found have always one
tachyonic direction that is a mixture of all the moduli. This
tachyonic direction is steep and leads to $\eta \sim O(1)$, so that
these dS critical points are neither stable nor suitable for
slow-roll inflation. For a large range of parameters there is only
this one tachyon and the other up to 13 directions are stable. There
are however also solutions where more directions are tachyonic
\cite{Danielsson:2010bc}\footnote{The models that are further
truncations to just three complex moduli have always only one
tachyonic direction.}. Due to the complexity of the potential we
were not able to find an analytic expression for the one seemingly
universal tachyonic direction. In the simplest model the tachyonic
direction seems to be determined by the root of an irreducible
polynomial of degree 19 so that there is no hope of getting an
analytic expression\footnote{We thank Mike Stillman for studying the
problem using the program Macaulay 2.}. It is not possible to tell
whether this tachyon is generic for geometric $\SU(3)$-structure
flux compactification with O6-planes or whether it is model
dependent. We would like to point out that the existence of the
tachyon is independent of the orientifold projection and therefore
the tachyon is present in two very different classes of models.
Nevertheless, the orbifold group in the models that have dS critical
points always contains $\mathbb{Z}_2 \times \mathbb{Z}_2$ so that
the tachyon might very well be related to this fact.

In a non-geometric compactification of an $\mathbb{Z}_2 \times
\mathbb{Z}_2$ orbifold there are new  terms in the superpotential
and those can lift the tachyon and lead to stable dS vacua
\cite{deCarlos:2009qm}. It is not known whether it is possible to
lift the tachyon in a geometric setup by adding further ingredients
like D$p$-branes, NS5-branes or KK-monopoles and it would be
interesting to investigate these possibilities.

\section{Flux and charge quantization}\label{quantization}
In addition to solving the supergravity equations of motion we have
to take into account flux and charge quantization which are `stringy
constraints'. Whether a solution still exists after imposing these
quantization conditions depends on how sensitive a solution is to a
variation of the flux parameters. For instance, the SUSY AdS
solutions of \cite{DeWolfe:2005uu,
Lust:2004ig,Koerber:2008rx,Caviezel:2008ik} are quite robust against
these variations since the flux parameters are not entirely fixed in
the solutions and there are still many fluxes that can be chosen
freely. This is in contrast with the de Sitter solutions, for which
it has been observed that the solutions correspond to small
``islands'' in parameter space \cite{Danielsson:2009ff,
Danielsson:2010bc}.

We will study the charge and flux quantization for the
explicit example of a dS extremum found from a 4D
approach in \cite{Caviezel:2008tf} and lifted to ten dimensions in
\cite{Danielsson:2010bc}. The compact space is the standard
orientifold projection of $\SU(2) \times \SU(2)/\mathbb{Z}_2 \times
\mathbb{Z}_2$ (see subsection \ref{sec:standarddetails}). The dS extremum
is also a dS extremum of $\SU(2) \times \SU(2)/\Delta(12)$
which means that all three K\"ahler moduli as well as three complex structure
moduli are equal: $k \equiv k_1 = k_2 =k_3$ and $\mathcal{Z}_2= \mathcal{Z}_3 =\mathcal{Z}_4$ (see subsection \ref{delta12stand}).

\subsection{Quantization of the dS solution on $\SU(2)\times \SU(2)$}

In appendix \ref{app:quant} we explain how we should in general
quantize fluxes and charges in the presence of a non-trivial
$H$-field. Let us now apply this to the dS solution of
\cite{Danielsson:2010bc}.

The bottom line of appendix \ref{app:quant} is that we should impose the quantization of $H$.
As for the RR-fluxes, since the even $H$-twisted homology is
\eq{
H^H_{\text{even}}(\SU(2)\times \SU(2),\mathbb{Z}) =
\mathbb{Z}_{n_H} \, , }
where the $\mathbb{Z}_{n_H}$
is spanned by the ordinary homology class of a point $[\{p\}]$, we need to impose
quantization of the Romans mass $\hat{F}_0$. As we can infer from
eq.~\eqref{quantmj} this implies the quantization of the charge of
the source $j$. In fact, the orientifold charge for an O6-plane
should be $-2$. However we can add D-branes on top of the O6-plane. Taking into account
that the Maldacena-N\'u\~nez no-go theorem \cite{Maldacena:2000mw}
requires the charge to be negative, it can thus be $-2,-1$ leading
to
\eq{ (n_H,n_{\hat{F}_0}) = (\pm 1,\pm 2),(\pm 2, \pm 1), (\pm 1, \pm 1) \,
. }
From the quantization of $\hat{F}_0$,
\eq{ \label{quantm} (2 \pi l_s) \hat{F}_0 =n_{\hat{F}_0} \, , }
we find using \eqref{fluxansatz} that
\eq{f_1 = \frac{e^{\Phi}}{2 \pi l_s} n_{\hat{F}_0} \, , }
or factoring out the K\"ahler modulus $k$, $f_i = \hat{f}_i k^{-1/2}$,
\eq{ \hat{f}_1 = e^{\Phi} \frac{k^{1/2}}{2\pi l_s} n_{\hat{F}_0} \, . }
At this point it seems that we can have an internal space with large radius $\frac{k^{1/2}}{2\pi l_s}$ and a small dilaton $e^{\Phi}$. From the quantization of $H$, eq.~\eqref{quantH}, we find
\eq{ C_{S^3}\frac{k}{(2\pi l_s)^2} \hat{h}=n_H \, , }
and for dS solution we have \cite{Danielsson:2010bc}
\eq{ \hat{h} = \frac{2(k^6 -3 \mathcal{F}_2^4)\hat{f}_5 + \sqrt{3}(k^6+\mathcal{F}_2^4)\hat{f}_6}{2 \mathcal{F}_2^3 k^{3/2}} = \lp \mathcal{Z}_2^3 -\frac{3}{\mathcal{Z}_2} \rp \hat{f}_5 + \frac{\sqrt{3}}{2} \lp \mathcal{Z}_2^3 + \frac{1}{\mathcal{Z}_2} \rp \hat{f}_6 \,,}
since $\mathcal{F}_2= \mathcal{F}_3=\mathcal{F}_4 = \frac{k^{3/2}}{\mathcal{Z}_2}$ as explained in subsection \ref{sec:standarddetails}.

\begin{figure}
\psfrag{f2f1}{$\scriptstyle f_2/f_1$}
\psfrag{fhsqr}{$\scriptstyle \hat{f}_1\hat{h}^{1/2}$}
\psfrag{hinvsqr}{$\scriptstyle \hat{h}^{-1/2}$}
\begin{center}
\subfigure[Plot of $\hat{h}^{-1/2}$ measuring the scale $\frac{k^{1/2}}{2\pi l_s}$.]{
\includegraphics[width=7cm]{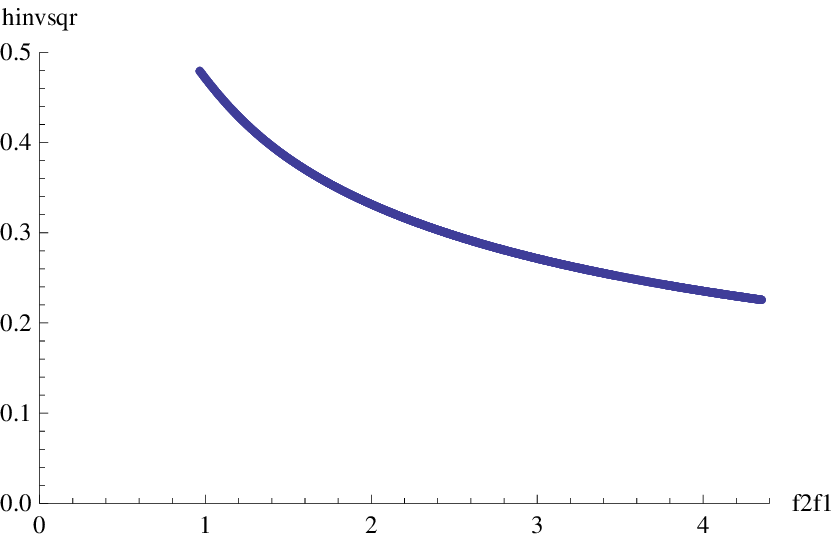}
}\hspace{0.2cm}
\subfigure[Plot of $\hat{f}_1\hat{h}^{1/2}$ measuring the dilaton $e^{\Phi}$.]{
\includegraphics[width=7cm]{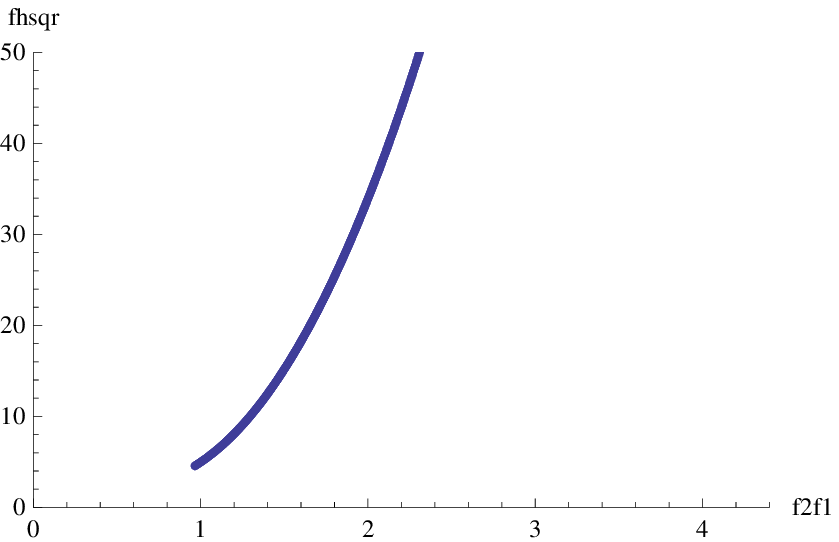}
}
\end{center}
\caption{Plots of radius and dilaton for the family of properly
quantized dS solutions on $\SU(2) \times \SU(2)$.}
\label{quantplots}
\end{figure}

For the supergravity regime we want to take the scale $k/(2\pi l_s)^2$
large, so that the scale independent $\hat{h}$ should be
small. However, this is impossible for the range of parameters that give dS solutions. To be more precise, in figure \ref{quantplots} we plot the
quantities
\eq{ \hat{h}^{-1/2} = (C_{S^3}/n_H )^{1/2} \frac{k^{1/2}}{2\pi l_s} \, , \qquad \hat{f}_1 \hat{h}^{1/2} = n_{\hat{F}_0}
(n_H/C_{S^3})^{1/2} e^{\Phi} \, , }
for the entire range of parameters that give dS solutions. We see the volume is always small and the string coupling is always large. Numerically, at $f_2/f_1 = 0.9648$ (the edge of the family of dS solutions)
$\hat{h}^{-1/2} \approx 0.4795$ and $\hat{f}_1\hat{h}^{1/2} \approx 4.554$, which
are respectively the maximum and minimum.
We conclude that flux quantization is impossible at tree-level in the supergravity
regime i.e.\ there are no properly quantized solutions with large internal volume and small string coupling.

\section{Discussion}\label{Discussion}
In this paper we have reviewed the current status of constructing
classical dS vacua in type IIA flux compactifications on
$\SU(3)$-structure manifolds with O6-planes. As summarized below, we
presented also several new results and pointed out important future
research directions.

We first explained in detail the 4D as well as the 10D approach that
have both been used in the literature. Then we classified
homogeneous group spaces that are consistent with an
$\SU(3)$-structure, extending the previous analyses
\cite{Grana:2006kf, Koerber:2008rx, Flauger:2008ad}. We discussed
orbifold and orientifold projections of these group spaces that lead
to an $\mathcal{N}=1$ supergravity in 4D. Here we paid special
attention to the abelian orbifold group $\mathbb{Z}_2 \times
\mathbb{Z}_2$ which is (presumably) the only abelian orbifold that
can lead to dS extrema. For the first time in the type IIA flux
compactification literature we discussed non-abelian orbifolds. We
were able to exclude several orbifold groups but also showed that
$\Delta(12)$ allows for dS extrema. We classified all group spaces
that are compatible with the $\mathbb{Z}_2 \times \mathbb{Z}_2$ or
$\Delta(12)$ orbifold groups and that lead to dS extrema.
Unfortunately, all the numerical dS extrema we found have at least
one tachyonic direction with $\eta \lesssim -2$ so that these
solutions are not stable and incompatible with slow-roll inflation.
This numerical analysis revealed that semi-simple groups almost
always lead to dS extrema (except for
$\frak{so}(3)\times\frak{so}(2,1)$ as it is not consistent with the
symmetries). Furthermore we found many examples amongst solvable
algebras of order 3 and one example for a solvable algebra of order
2. Finally there were some examples for algebras that are a
semi-direct product of a 3-dimensional solvable algebra with a
3-dimensional simple algebra. We could not find dS extrema for any
of the nilmanifolds. This is somewhat surprising since nilmanifolds
are everywhere negatively curved and negative curvature is a
necessary condition for a dS solution. In contrast,
$\SU(2)\times\SU(2)$ is only negatively curved in ``a corner of its
moduli space'', but that corner harbors dS extrema.

Finally, we discussed the flux and charge quantization. This has
been neglected so far in the literature since the internal spaces
have generically a rather complicated geometry which aggravates the
analysis. We worked out the flux quantization explicitly for the
example of $\SU(2) \times \SU(2)/\mathbb{Z}_2 \times \mathbb{Z}_2$
\cite{Caviezel:2008tf}, whose 10D origin is best understood
\cite{Danielsson:2010bc}, and found that the small range of
parameters that lead to dS extrema are not compatible with flux and
charge quantization in the supergravity regime. This presents a
great new challenge for all the constructions in the literature.

Throughout the paper we pointed out many important questions:
\begin{itemize}
\item Is it possible to find stable dS vacua or is the tachyonic direction
universal?\footnote{There does not seem to be a simple structure in
the tachyonic modes that would allow us to understand why all
solutions sofar are unstable. Though some progress has been made to
understand the presence of tachyonic modes for de Sitter solutions
in four-dimensional supergravity theories \cite{Covi:2008ea,
GomezReino:2008bi, Borghese:2010ei}, it is not clear yet how the
models with a known higher-dimensional origin, like ours, fit into
that discussion.}
\item Does the scalar potential allow for slow-roll? \item Is flux and
charge quantization generically incompatible with the existence of
dS extrema or is this only the case for our explicit $\SU(2) \times
\SU(2)/\mathbb{Z}_2 \times \mathbb{Z}_2$ example?
\item Can one exclude all non-abelian orbifold groups except $\Delta(12)$ or are
other non-abelian orbifold groups compatible with dS extrema?
\item Is it possible to include the back-reaction of the O6-planes and how
does this affect the solutions \cite{Douglas:2010rt,
Blaback:2010sj,Blaback:2011}?
\item Is it possible to study compactifications on more general
geometries like for example the smooth, compact toric varieties
discussed in \cite{Larfors:2010wb}?
\end{itemize}
We hope that this paper provides the technical tools necessary to
embark on answering some of these questions.

\section*{Acknowledgements}
We like to thank David Andriot, Andres Collinucci, Raphael Flauger,
Daniel Junghans, Liam McAllister, Susanne Reffert, Daniel Robbins,
Mike Stillman, Henry Tye and Marco Zagermann for useful discussions.
U.D.\ is supported by the Swedish Research Council (VR) and U.D.\
and T.V.R.\ are supported by the G\"{o}ran Gustafsson Foundation.
P.K.\ is a Postdoctoral Fellow of the FWO -- Vlaanderen. The work of
P.K.\ is further supported in part by the FWO -- Vlaanderen project
G.0651.11 and in part by the Federal Office for Scientific,
Technical and Cultural Affairs through the `Interuniversity
Attraction Poles Programme Belgian Science Policy' P6/11-P. The work
of S.S.H.\ and G.S.\ was supported in part by NSF CAREER Award No.\
PHY-0348093, DOE grant DE-FG-02-95ER40896, a Research Innovation
Award and a Cottrell Scholar Award from Research Corporation, a
Vilas Associate Award from the University of Wisconsin, and a John
Simon Guggenheim Memorial Foundation Fellowship. T.W.\ is supported
by the Alfred P. Sloan Foundation and by the NSF under grant
PHY-0757868. T.V.R. thanks the University of Wisconsin, Madison for
hospitality. G.S. and T.W. thank the Hong Kong Institute for
Advanced Study for hospitality.

\appendix

\section{Form conventions for reduction to 4D}
\label{sec:Conventions}

Consider type IIA string theory on an $\SU(3)$-structure manifold
$M_6$, equipped with a $\mathbb{Z}_2$ orientifold action which
includes an anti-holomorphic involution $\sigma$. The forms on $M_6$
then split into even and odd parts, depending upon the behavior of
each class under $\sigma$. We will take the following basis of
representative real forms\footnote{The existence of everywhere non-vanishing one-forms would
imply that the structure group is a strict subgroup of $\SU(3)$ like
for example $\SU(2)$ or the trivial group. We do not consider such
cases here.}:
\begin{itemize}
\item The zero-form 1,
\item a set of odd two-forms $Y^{(2-)}_i$, $i=1,\ldots,h^{1,1}_-$,
\item a set of even two-forms $Y^{(2+)}_\a$, $\a=1,\ldots,h^{1,1}_+$,
\item a set of even four-forms $Y^{(4+)i}$, $i=1,\ldots,h^{1,1}_-$,
\item a set of odd four-forms $Y^{(4-)\a}$, $\a=1,\ldots,h^{1,1}_+$,
\item a six form $Y^{(6-)}$, odd under $\sigma$,
\item a set of even three-forms $Y^{(3+)}_K$, $K=1,\ldots,h^{2,1}+1$,
\item and a set of odd three-forms $Y^{(3-)K}$, $K=1,\ldots,h^{2,1}+1$.
\end{itemize}
It turns out that we can always choose the $Y^{(3+)}_K$ and $Y^{(3-)K}$ to form a symplectic basis such that the only non-vanishing intersections are
\be
\int Y^{(3+)}_K\w Y^{(3-)J}=\delta_K^J.
\ee
Furthermore, we define the triple intersecting numbers
\be
\k_{ijk} = \int Y^{(2-)}_i \w Y^{(2-)}_j \w Y^{(2-)}_k, \qquad \hat{\k}_{i \a\b} = \int Y^{(2-)}_i \w Y^{(2+)}_\a \w Y^{(2+)}_\b ,
\ee
and take the even degree forms to satisfy
\be
\int Y^{(6-)} =1,\qquad \int Y^{(2-)}_i \w Y^{(4+)j} = \delta_i^j,\qquad \int Y^{(2+)}_\a \w Y^{(4-)\b} = \delta_\a^\b.
\ee

\section{Half-flat manifolds}\label{halfflat}
A six-dimensional $\SU(3)$-structure manifold can be characterized
by a globally defined real two-form $J$ and a complex decomposable
three-form $\Omega=\Omega_R+i\Omega_I$, satisfying a compatibility
and a normalization condition
\eq{ \Omega \wedge J=0 \, , \qquad
\Omega\wedge\Omega^*=(4i/3) \, J\wedge J\wedge J = 8i
\label{normcond} \, \text{vol}_6 \, . }
From the real part of the
three-form we can build an almost complex structure for which $J$ is
of type $(1,1)$ and $\Omega$ is of type $(3,0)$. It is given by
\begin{equation}\label{complexstructure}
I^l_{\,\,k}= c \, \varepsilon^{m_1m_2\ldots m_5l}(\Omega_R)_{km_1m_2}(\Omega_R)_{m_3m_4m_5}\,,
\end{equation}
where $\varepsilon$ is the Levi-Civita symbol, and the real scalar
$c$ is such that $I$ is properly normalized: $\tfrac16 \text{tr}(I^2) = -1$. The metric
then follows via
\begin{equation}\label{metricfromI}
g_{mn}=-I^l_{\,\,m}J_{ln}\,.
\end{equation}

The torsion classes $W_1,\ldots, W_5$ correspond to the expansion of
the exterior derivatives of $J$ and $\Omega$ in terms of
$\SU(3)$-representations \cite{Chiossi:2002tw}. An
$\SU(3)$-structure (that is not also an $\SU(2)$-structure) has no
nowhere-vanishing one-forms. If we restrict to left-invariant
torsion classes on an homogeneous manifold, this implies that the
torsion classes $W_4$ and $W_5$ have to vanish.  In the absence of
D-terms the $\SU(3)$-structure manifold is a so-called {\em
half-flat manifold} \cite{Ihl:2007ah}.  In our conventions this
corresponds to $W_1, W_2$ real. We will often restrict to this case
although it would be interesting to study the more general setup.
For a half-flat manifold we have \subeq{\label{tcl}
\begin{align}
&\d J = \frac{3}{2}W_1\Omega_R + W_3\,,\\
&\d\Omega_R = 0 \, , \\
&\d\Omega_I = W_1 J\wedge J + W_2\wedge J\, ,
\end{align}
}
where $W_1$ is a real scalar, $W_2$ a real primitive $(1,1)$-form and $W_3$ a real primitive $(1,2)+(2,1)$-form. This means that
\subeq{\al{
\hspace{1cm} & W_2\wedge J\wedge J=0\, , && W_3 \wedge J = 0 \, , \hspace{1cm} \\
& W_2\wedge \Omega=0 \, ,  &&  W_3 \wedge \Omega = 0 \, . }}
Furthermore, we find that under the Hodge star, defined from the metric \eqref{metricfromI},
\eq{\label{hodgeprop}
\star_6\Omega=-i\Omega \, , \qquad \star_6 J=\tfrac{1}{2}\,
J\wedge J  \, , \qquad \star_6 W_2=-J\wedge W_2 . }

The Ricci tensor can be expressed in terms of the torsion classes
\cite{bedulli-2007-4,Ali:2006gd}. For that we use that any real
symmetric two-tensor $T_{ij}$ splits as follows in representations
of $\SU(3)$ \eq{\label{symdecomp} T_{ij} = \frac{s(T_{ij})}{6}
g_{ij} + T_{ij}^+ + T_{ij}^- \, . } $s(T_{ij})$ is the trace, an
$\SU(3)$-invariant, and $T^+_{ij}$ and $T^-_{ij}$ transform
respectively as $\bf{8}$ and $\bf{6}+\bf{\bar{6}}$. The latter are
traceless and have respectively index structure (1,1) and
(2,0)+(0,2) \subeq{\al{
& T_{ij}^+ g^{ij} = 0 \, , \qquad I^i{}_k T^+_{ij} I^j{}_l = T^+_{kl} \, , \\
& T_{ij}^- g^{ij} = 0 \, , \qquad I^i{}_k T^-_{ij} I^j{}_l =
-T^-_{kl} \, . }}
We can associate a primitive real (1,1)-form and a complex primitive (2,1)-form to respectively $T_{ij}^+$ and $T_{ij}^-$ \subeq{\label{symdecomp2}\al{
\label{symtwoformpart} & P_2(T_{ij}) = \frac{1}{2} J^k{}_i T^+_{kj} \, \d x^{i} \wedge \d x^j   \, , \\
\label{symthreeformpart} & P_3(T_{ij}) = \frac{1}{2} T^-_{il}
\Omega^l{}_{jk} \, \d x^{i} \wedge \d x^j \wedge \d x^{k} \, . }}
Using this it is shown in \cite{bedulli-2007-4} that the Ricci
tensor can be expressed as follows \subeq{\al{
s(R_{ij}) & = \frac{15}{2} (W_1)^2 - \frac{1}{2} (W_2)^2 -\frac{1}{2} (W_3)^2 \, , \\
P_2(R_{ij}) & = - \frac{1}{4} \star (W_2 \wedge W_2)  - \frac{1}{2} \star_6 d \star_6 \left(W_3 - \frac{1}{2} W_1 \Omega_R\right) \, , \label{ricci2} \\
P_3(R_{ij}) & = 2 \, W_1 W_3|_{(2,1)} + 2 \, \d W_2|_{(2,1)} -
\frac{1}{4} Q_1(W_3,W_3) \, , \label{ricci3} }}
with
\eq{ Q_1(W_3,W_3) = \left(\Omega^{ijk} \iota_j \iota_i W_3 \wedge
\iota_k W_3\right)_{(2,1)} \, , }
and where in the right-hand side of eqs.~\eqref{ricci2}-\eqref{ricci3} the projection onto the primitive part is understood.

It is also useful to define the following quantity
\begin{equation}
Q_2(\hat{W}_3,\hat{W}_3) = \left.\left(\frac{1}{2} \hat{W}_{3\,imn}
\hat{W}_3{}^{pmn} \Omega_{pjk} \, \d x^i \wedge \d x^j \wedge \d
x^k\right)\right|_{(2,1)}\,.
\end{equation}

\section{10D equations of motion for the universal ansatz}\label{ap:universaleoms}
If we plug the universal ansatz (\ref{fluxansatz}) into the type IIA
SUGRA equations of motion and assume the constraint equations
(\ref{tclprop}) we find, after a lengthy calculation, the following
algebraic equations
 \allowdisplaybreaks \subeq{\al{ & \left(
\frac{3}{2} f_2 W_1 - \frac{1}{4} f_3 w_2 + f_1 f_7 + j_1 \right)
\Omega_R +
\left( f_2 w_3 + f_3 d_1 + f_1 f_8 + j_2 \right) \hat{W_3} =0 \qquad (\text{Bianchi }\hat{F}_2) \, , \\
& \left( 3 f_4 W_1 + \frac{1}{4} f_5 w_2 -f_6 f_7 \right)  \Omega_R
+\left( 2 f_4 w_3 - f_5 d_1 - f_6 f_8 \right) \hat{W}_3 = 0
\qquad (\text{eom }\hat{F}_4) \, , \\
& \left( e^{-2\Phi} f_7 W_1 + e^{-2\Phi} \frac{f_8 w_3}{6} - \frac{1}{2} f_1 f_2 - 2 f_2 f_4 - \frac{1}{6} f_3 f_5 - f_4 f_6 \right) J \wedge J \nonumber \\
& + \left( e^{-2 \Phi} f_7 w_2 - e^{-2 \Phi} f_8 d_1 + f_1 f_3 - 2 f_3 f_4 + f_2 f_5 + f_3 f_5 d_2 - f_5 f_6 \right) J \wedge \hat{W}_2=0 \quad (\text{eom }H) \, , \\
& R_4 = - \frac{15}{2} (W_1)^2  + \frac{1}{2}\left[(w_2)^2+(w_3)^2+(f_8)^2 \right] + 2 f_7^2 + 2 e^{\Phi} j_1  \quad (\text{dilaton eom}) \, , \\
& R_4 + e^{2 \Phi} \left[ (f_1)^2 + 3(f_2)^2 + (f_3)^2 + 12 (f_4)^2 +(f_5)^2 + (f_6)^2\right] + 4 e^{\Phi} j_1 = 0 \nonumber\\
& \qquad\qquad\qquad\qquad\qquad\qquad\qquad\qquad\qquad (\text{external part Einstein eq.}) \, , \\
& -2 (f_7)^2 - \frac{1}{2} (f_8)^2 + \frac{e^{2\Phi}}{4} \left[ 5 (f_1)^2 + 9 (f_2)^2 +12 (f_4)^2 - (f_6)^2 + 3 (f_3)^2 + (f_5)^2\right] +3 e^{\Phi} j_1 = 0 \nonumber \\
& \qquad\qquad\qquad\qquad\qquad\qquad\qquad\qquad\qquad (\text{trace Einstein/dilaton eom}) \, , \\
& \frac{d_2 (w_2)^2}{4} -\frac{d_1 w_3}{2} - \frac{W_1 w_2}{4} + \frac{(f_8)^2 d_4}{2} + \frac{e^{2\Phi}}{4} \left[((f_5)^2-(f_3)^2) d_2 + 4 f_2 f_3 + 8 f_4 f_5 \right] = 0 \nonumber \\
& \qquad\qquad\qquad\qquad\qquad\qquad\qquad\qquad\qquad (\text{Einstein eq., two-form part}) \, , \\
& 2 (W_1 w_3 + d_1 w_2 - e^{\Phi} j_2 - 2 f_7 f_8) \hat{W}_3|_{(2,1)}  - \frac{1}{4}  \left((w_3)^2 Q_1(\hat{W}_3,\hat{W}_3) + (f_8)^2 Q_2(\hat{W}_3,\hat{W}_3)\right) = 0 \nonumber \\
& \qquad\qquad\qquad\qquad\qquad\qquad\qquad\qquad\qquad
(\text{Einstein eq., three-form part}) \, . }}

\section{Flux/charge quantization with non-trivial $H$-field}
\label{app:quant}

\subsection{General discussion}

In this appendix we study in some detail the quantization of the
NSNS flux $H$, the RR-fluxes and the charge of the orientifold
planes. The easiest is the quantization condition for $H$, which
just reads \eq{ \label{Hquant} \frac{1}{(2\pi l_s)^2}
\int_{\Sigma^i} H = n^i_H \, , } with $n^i_H$ integer for all cycles
$\Sigma^i$ that are non-trivial in homology. Here, $l_s =
\alpha'{}^{1/2}$ is the string length. For the charge of the
orientifold plane we have to take into account that the source $j$
entering the Bianchi identities, is in fact given by \eq{j =
\sum_{\text{O}p} (2 \kappa_{10}^2) T_p m_p \, j_{\text{PD}} \, , }
where $2 \kappa_{10}^2=(2\pi)^7 l_s^8$, $T_p=(2\pi)^{-p} l_s^{-p-1}$
is the D-brane tension, $m_p = -2^{p-5}$ the proportionality between
the O-plane and the D-brane tension, and $j_{\text{PD}}$ is the
actual Poincar\'e dual to the submanifold (see
\cite{deformations,Koerber:2007hd,Koerber:2010bx} for exact
definition and conventions for $j$) that the orientifold wraps. For
each O-plane wrapping a non-trivial cycle $\Sigma^i$, we find,
plugging in all the factors, for the O-plane charge \eq{ (2\pi
l_s)^{p-7} \int_{\Sigma^i} j = m_p = -2^{p-5} \, .} The more subtle
part is the quantization of the RR-fluxes, for which, in the
presence of non-trivial $H$-flux, one should in fact use $H$-twisted
K-theory \cite{Ktheory1,Ktheory2}. However, we will use here the
result of \cite{twistedhom}, where it is claimed that for the case
of a simply-connected six manifold $H$-twisted K-theory is
isomorphic to $H$-twisted homology. We will study the quantization
for the dS solution on $\SU(2) \times \SU(2)$, which is indeed
simply-connected, and use the approach of twisted homology.

In order to define the twisted boundary operator, let us first consider what constitutes a consistent D-brane \cite{evslinKtheoryreview,twistedcohom}.
Consider the D-brane's Chern-Simons action,
\eq{
\label{CSterm}
S_{\text{CS}} = T_p \int_{\Sigma} C \wedge e^{\mathcal{F}} \, ,
}
with $\mathcal{F}$ the D-brane world-volume flux, satisfying
\eq{
\label{dF}
\d \mathcal{F} = H|_\Sigma \, .
}
This action is only invariant under the gauge transformation of the RR-fields, $\delta C = \d_H \Lambda$,
if $\Sigma$ has no boundary. So a consistent D-brane should wrap a submanifold without boundary, a cycle.
On the other hand, if the D-brane wraps a cycle which is itself a boundary (of some $\Gamma$) then its charge is trivial because it
can shrink to zero by sweeping out $\Gamma$.

Now, if we introduce non-trivial $H$-field this story is changed in two ways \cite{twistedcohom}. First of all, as will become clear in a moment, we
are forced to consider networks of several D-branes wrapping submanifolds $\Sigma_k$ of different dimensions $k$ (but still of the same parity). Secondly, we need to relax eq.~\eqref{dF} to allow for magnetic monopoles of the D-brane world-volume field
\eq{
\d \mathcal{F} = H|_{\Sigma_k} + (2\pi l_s)^2 \delta^3(\mathcal{C}_{k-3}) \, ,
}
where $\mathcal{C}_{k-3}$ is codimension three within $\Sigma_k$. In this way we allow the cohomology class of $H|_{\Sigma_k}$
to be non-trivial. More precisely, the submanifold $\mathcal{C}_{k-3}$ is homologous to the Poincar\'e dual of $H|_{\Sigma_k}$
within $\Sigma_k$. Revisiting the argument of gauge invariance of the Chern-Simons action eq.~\eqref{CSterm} we find that
$\mathcal{C}_{k-3}$ should be considered in the same way as the boundary of $\Sigma_k$. Namely, for a consistent network of D-branes the sum of the
ordinary boundaries and these Poincar\'e duals should vanish. Since the ordinary boundary is of codimension one and the Poincar\'e dual of codimension three, the necessity of considering networks of D-branes of different dimensions becomes clear. Likewise, there is a new way for a D-brane to decay and thus have trivial charge. Indeed, if it wraps a submanifold corresponding to the magnetic monopole of $\mathcal{F}$ it can dissolve in the larger D-brane. So we are naturally led to consider the homology of the twisted boundary operator. In \cite{twistedhom} this operator is defined as follows
\eq{
\partial_H = \partial + H \cap \, ,
}
where $\partial$ is the ordinary boundary operator and $H \cap\,$ produces the codimension-three
Poincar\'e dual of $H|_{\Sigma_k}$ within $\Sigma_k$.\footnote{More accurately, $\mathcal{C}_{k-3}$ is
the Poincar\'e dual of $H|_\Sigma - \d \mathcal{F}$. In \cite{twistedhom} the world-volume gauge field $\mathcal{F}$
is put to zero. This, however, does not affect the homology of the twisted boundary operator as it only depends on the
cohomology class of $H$. So we will also put $\mathcal{F}=0$ in the following.}

So let us now discuss quantization of the RR-fluxes using twisted homology \cite{twistedcohom}. For each consistent network of D-branes, we want the contribution of the Chern-Simons action to the path integral to be well-defined. Let us write the Chern-Simons term in a manifestly gauge-invariant way as follows.
For an even/odd network $\sum_{k \in \text{E/O}} (\Sigma_k,\mathcal{F}_{\Sigma_k})$ we choose a fixed reference
network $\sum_{k \in \text{E/O}} (\Sigma^0_{k},\mathcal{F}_{\Sigma^0_{k}})$ in the same twisted homology class, such that there exists an
odd/even network $\sum_{k \in \text{O/E}} (\tilde{\Gamma}_k,\mathcal{F}_{\tilde{\Gamma}_k})$ satisfying
\eq{
\partial_H \left( \sum_{k \in \text{O/E}} (\tilde{\Gamma}_k,\mathcal{F}_{\tilde{\Gamma}_k})\right) = \sum_{k \in \text{E/O}} (\Sigma_k,\mathcal{F}_{\Sigma_k}) - \sum_{k \in \text{E/O}} (\Sigma^0_k,\mathcal{F}_{\Sigma^0_k}) \, .
}
We find then:
\eq{ S_{\text{CS}} = S_{\text{CS}}^0 + \sum_{\tilde{\Gamma}_k} T_{k-1} \int_{\tilde{\Gamma}_k}
F \wedge e^{\mathcal{F}_{\tilde{\Gamma}_k}} \, , }
where the contribution $S_{\text{CS}}^0$ from the reference chain is just
a constant. We want the contribution of the Chern-Simons action to
the path integral,
\eq{ \exp (2\pi i S_{\text{CS}}) \, , }
to be independent of the choice of network $\sum_{k \in \text{O/E}}
(\tilde{\Gamma}_k,\mathcal{F}_{\tilde{\Gamma}_k})$. Two such networks would
differ by an odd/even $H$-twisted cycle $\sum_{k \in \text{O/E}}
(\Gamma_k,\mathcal{F}_{\Gamma_k})$. So in the end we find the quantization
condition
\eq{ \label{intF} \sum_k \frac{1}{(2 \pi l_s)^{k-1}} \int_{\Gamma_k} F \wedge e^{\mathcal{F}_{\Gamma_k}} \in \mathbb{Z} }
for every $H$-twisted cycle $\sum_{k \in
\text{O/E}}(\Gamma_k,\mathcal{F}_{\Gamma_k})$. In words, we need the integral
of the RR-fluxes $F$ to be an integer upon integrating over a
non-trivial cycle in $H$-twisted homology.

\subsection{The $H$-twisted homology of $\SU(2) \times \SU(2)$}

As an example, let us calculate the $H$-twisted homology of $\SU(2)
\times \SU(2)$. Let us take
\eq{ H = \alpha (e^{123} - e^{456}) \, ,}
with $\alpha$ some proportionality constant. The non-trivial
three-cycles are the two $S^3$s $\Sigma_1$ and $\Sigma_2$. Imposing
the quantization of $H$, eq.~\eqref{Hquant}, for these two cycles,
we find
\eq{ \label{quantH} \frac{1}{(2\pi l_s)^2} \int_{\Sigma_1} H
= - \frac{1}{(2\pi l_s)^2} \int_{\Sigma_2} H = n_H \, , }
which fixes $\alpha = \frac{n_H(2 \pi l_s)^2}{C_{S^3}}$ with $C_{S^3} =
\int_{\Sigma_1} e^{123} = 16 \pi^2$.

One can show that $H$-twisted homology is dual to the cohomology of the $H$-twisted exterior derivative \cite{twistedcohom}
\eq{
\d_H = \d + H \wedge \, .
}
We can therefore try to calculate
the $H$-twisted cohomology first. We know that for a coset manifold, the ordinary cohomology is isomorphic to the
cohomology of left-invariant forms (see e.g.~\cite{cosetreview1}). Let us assume that this also applies to $H$-twisted cohomology.
One can construct a proof of this using the spectral sequence method outlined below. Calculating the $\d_H$-cohomology of left-invariant forms one finds
in a straightforward way:
\eq{
H^H_{\text{E}}(\SU(2)\times\SU(2),\mathbb{R})=H^H_{\text{O}}(\SU(2) \times\SU(2),\mathbb{R})=\bf{1} \, ,
}
where $E/O$ indicate the even respectively odd cohomology.\footnote{Just as for the D-branes we have to take
sums of forms of different dimensions, but of the same parity. So instead of finding a cohomology for every dimension of forms, we just have
an even and an odd cohomology.}
However, this is a bit quick, because we need the integer cohomology instead of the real one.

So let us instead calculated the $H$-twisted homology directly. We use the spectral sequence method explained in more detail in \cite{twistedhom}.
There the twisted homology is calculated by a series of approximations $E^i_{\text{E}}$ and $E^i_{\text{O}}$ that eventually converges
to the correct answer. At each step one defines a differential operator:
\eq{
\d_{i}: E^i_{\text{E/O}} \longrightarrow E^i_{\text{O/E}} \, ,
}
and the next step in the series is the cohomology of this operator:
\eq{
E^{i+1}_{\text{E/O}} = \frac{\text{Ker}(\d_{i}:E^i_{\text{E/O}}\rightarrow E^i_{\text{O/E}})}{\text{Im}(\d_{i}:E^i_{\text{O/E}}\rightarrow E^i_{\text{E/O}})} \, .
}
Now, we take $E^0_{\text{E/O}}$ the sets of even and odd chains and
\eq{
\d_1 = \partial \, , \qquad \d_2 = H \cap \, .
}
So $E^1$ is in fact the untwisted homology, while for a six-dimensional simply-connected orientable manifold one can show that $E^2$ has already
converged to the correct answer.

Let us now apply this to $\SU(2)\times \SU(2)$. The untwisted
homology is given by \eq{\spl{
H_{\text{E}}(\SU(2)\times\SU(2),\mathbb{Z}) & = \text{span}_{\mathbb{Z}}([\{p\}],[\Sigma_1] \times [\Sigma_2]) \, \\
H_{\text{O}}(\SU(2)\times\SU(2),\mathbb{Z}) & = \text{span}_{\mathbb{Z}}([\Sigma_1],[\Sigma_2]) \, .
}}
Let us now consider the action of $\d_2 = H \cap$:
\eq{\spl{
& H \cap [\{p\}] = 0 \, , \\
& H \cap [\Sigma_1] = - H \cap [\Sigma_2] = n_H [\{p\}] \\
& H \cap ([\Sigma_1] \times [\Sigma_2]) = (H \cap [\Sigma_1])\times [\Sigma_2] - [\Sigma_1] \times (H \cap [\Sigma_2]) = n_H([\Sigma_1] + [\Sigma_2]) \, .
}}
So the closed cycles are spanned by $[\{p\}]$ and $[\Sigma_1] + [\Sigma_2]$. On the other we find that the exact cycles are spanned by $n_H [\{p\}]$ and $n_H([\Sigma_1] + [\Sigma_2])$.
It follows that the even and odd $H$-twisted homology is given by
\eq{
H^H_{\text{E}}(\SU(2)\times \SU(2),\mathbb{Z}) = \mathbb{Z}_{n_H} \, , \qquad
H^H_{\text{O}}(\SU(2)\times \SU(2),\mathbb{Z}) = \mathbb{Z}_{n_H} \, .
}
So we find that there is discrete torsion, which of course we could not see by calculating the {\em real} cohomology.
As an example, in contrast to ordinary homology in {\em real} $H$-twisted homology a point is now not a non-trivial
cycle anymore, but is a generalized boundary. In the dual language the volume-form is trivial because it is exact.
So it would seem we do not need to quantize the Romans mass $\hat{F}_0$ anymore, but
\eq{\label{quantmj}\spl{
n_{\hat{F}_0} = (2 \pi l_s) \hat{F}_0 & = (2 \pi l_s) \frac{1}{(C_{S^3})^2}\int_M \hat{F}_0 e^{123456} \\
              & = \frac{1}{n_H} \frac{1}{2\pi l_s}\int_M \hat{F}_0 \, H \wedge \frac{1}{2C_{S^3}}(e^{123} + e^{456}) \\
              & = -\frac{1}{n_H} \, \frac{1}{2\pi l_s} \int_{\Sigma_1} j = \frac{1}{n_H} \, \frac{1}{2\pi l_s} \int_{\Sigma_2} j\\
              & = - \frac{1}{n_H} n_j \, ,
}}
where we used the Bianchi identity $\d \hat{F}_2 +\hat{F}_0 H = -j$. So the quantization of $j$ would imply
the quantization of $(2 \pi l_s) \hat{F}_0$ in units of $1/n_H$. But in fact the integer homology is not trivial but $\mathbb{Z}_{n_H}$, which
indeed implies the quantization of $(2 \pi l_s) \hat{F}_0$ as an integer.

\bibliography{desitter}

\bibliographystyle{utphysmodb}

\end{document}